\documentclass[reprint,aip,graphicx]{revtex4}
\usepackage{graphicx}
\usepackage{color}
\usepackage{ulem}
\usepackage{amsmath}

\begin{document}
\title{Microscopic Theory for the Dynamics of Unentangled and Entangled Polymer Melts}
\author{M. G. Guenza\footnote{ Electronic mail: mguenza@uoregon.edu}}
\affiliation{Department of Chemistry and Biochemistry, and Institute of Theoretical Science, University of Oregon, Eugene, Oregon 97403}
\date{\today}

\newpage
\begin{abstract}
 
The Langevin Equation for cooperative dynamics represents the dynamics of polymer melts with chains of increasing degree of polymerization, covering the full range of  behavior from the unentangled to the entangled regime. This equation describes the motion of a group of interpenetrating polymers that are interacting through an effective potential resulting from the many-body coupling of the inter polymer potential inside the macromolecular liquid.
The confinement of the dynamics due to the presence of entanglements is accounted for by an effective inter-monomer potential which is zero until the distance between two monomers belonging to different chains reaches a characteristic value, $d$. At that distance a constraint is applied through an effective hard-core repulsion that represents the effect of entanglements in the slowing down of the relative diffusion of the monomers. As the time evolves the constraint relaxes due to the chain interdiffusion.
The same potential acts on both unentangled and entangled polymer chains, but short chains are not affected as they relax faster than they experience the presence of the potential.
 
The formalism is a Langevin equation approach, consistent with the traditional Rouse model of unentangled polymer chains, but includes the effects of the chemical nature of the polymer in the form of local flexibility, the finite size of the molecule, the correlated dynamics due to intermolecular many-chain affects, and the possible presence of entanglements between chains. Theoretical predictions for the characteristic scaling  exponents of dynamical quantities are in agreement with experimental ones for both unentangled and entangled chains. The theory accounts for the experimentally observed cooperative motion of polymer melts and the related sub-diffusive dynamics of the center-of-mass, the crossover from unrestrained-local to restrained-global dynamics for entangled samples, and the change in the dynamics as a function of the chain length. The comparison with Neutron Spin Echo experiments of polyethylene melts  shows quantitative agreement with  the time dependent incoherent scattering function for a variety of samples in both the unentangled and the entangled regimes.

\end{abstract}

\maketitle

\section{Introduction}
Entangled polymer melts display unique dynamical properties as the time scales of diffusion and viscosity change several orders of magnitude with increasing degree of polymerization. More specifically, the scaling exponents that define the dynamics, diffusion and viscosity, as a function of the degree of polymerization, $N$, are different for short and for long chains. In long chains the dynamics is dominated by the presence of entanglements, transient points of contact between chains, due to the impossibility of chains to cross through each other.\cite{Colby,SQW,Liky} Such entanglements limit chain diffusion and, for a large number of entanglements per chain, experiment shows that diffusion occurs mainly along the direction of the curvature of the chain in a motion reminiscent of the slithering of a snake, i.e. the so-called ``reptation" dynamics.\cite{Chu,Granick} In liquids of short chains, however, the number of entanglements per chain is small, chain diffusion is fast, and the entanglements are short lived, so that their presence does not affect the motion of the chain. 

The Rouse model accurately represents the dynamics of melts of short chains through a Langevin approach that describes the time evolution of the coordinates of the monomers in a macromolecular chain. However the model does not account for the constraints due to the presence of entanglements as it represents the polymers as phantom chains free of crossing each other.  The chain in the Rouse model is, interestingly, fully flexible and infinitely long, to the point that the approximation of a continuum description becomes correct, but not entangled.\cite{doiedw} The implementation of the Rouse model to treat chains of finite length and with local semi-flexibility improves the agreement with experimental data.\cite{esterpaper,guenzareview} Including the intermolecular interactions and the dynamical slowing down of the center-of-mass in the Rouse model provides quantitative agreement for the sub-diffusive dynamics of polymer center-of-mass mean-square displacement.\cite{manychains,marina,marina1,marina2} 

The dynamics of long chains in a melt is traditionally described by the``reptation" model, where the fluid relaxation is due to the diffusion of a chain outside of the effective ``tube" formed by the presence of entanglements. The phenomenological reptation model was first described by de Gennes and subsequently formalized by Doi and Edwards.\cite{deGennes,doiedw} 
In an effort to improve the agreement between theory and experiment, several extensions to the reptation model have been developed over the following years.
It was observed that the original reptation model works fairly well in the "fully-entangled" regime, where the linear chains have a high degree of polymerization and the constraint of entanglements is long lasting.  However, it  provides limited  agreement with experiments for weakly entangled chains where it is difficult to define an effective ``tube". In that regime the original reptation model has to be modified to account for mechanisms of relaxation other then escape from the tube constraint, which are found to greatly improve the agreement of the theory with experiments.  Contour length fluctuations, constraint release, and tube dilation have been introduced with the motivation that the "tube" is formed by the chains surrounding the "tagged" chain and, if the polymers are short enough, both the interchain contacts and the effective tube evolve in time because the molecules diffuse away from each other.\cite{smith,MLL,wata}

A detailed discussion of the different implementations of the reptation model can be found in recent review articles.\cite{wata, Lodge, McLeish,Likhtman} In general it is observed that when these models are adopted, the agreement between the reptation theory and the experiments is greatly improved. However, it is also seen that models optimized to reproduce, for example, scattering data could hardly agree with data  measured otherwise, such as in dynamical mechanical experiments, or in simulations.\cite{Grest}  Furthermore, recent experiments in the non-linear dynamical regime, which have been recently confirmed by simulations, have suggested that the reptation model incorrectly explains the physics of the observed stress overshoot during high-rate startup shear.\cite{overshoot} 

A recent reptation-like approach has been presented by Likhtman and coworkers, which involves a simple description of entanglements as harmonic springs connected to a static support, mimicking the confinement of reptation.\cite{Likhtman} This so-called "slip-link" model describes the constrained dynamics of the entangled chains and their anisotropic chain diffusion. The model is an extension of Rubinstein's original approach,\cite{Rubinstein} and has proven quite successful in describing numerous aspects of entangled polymer dynamics. The approach is still phenomenological. It has been further extended by dePablo and others to include springs between pairs of interacting chains and thus ensure momentum conservation.\cite{depablo} An earlier version of the slip-link model was presented by Schiber and coworkers.\cite{Schiber,Schiber1} Another slip-link model of more recent construction was implemented in the New Algorithm for Polymeric Liquids Entangled and Strained (NAPLES), by Marrucci, Watanabe, and coworkers.\cite{Marrucci,Marrucci1}

While "reptation-like" approaches in general have been successful in predicting characteristic scaling exponents that agree with experimentally observed physical behavior, these models are limited when it is necessary to establish a direct connection between the chemical structure of the polymer, at defined thermodynamic conditions, and the measured dynamics.\cite{grest} Quantities such as the entanglement degree of polymerization, the tube diameter, and the plateau modulus are not unequivocally defined on the basis of the polymeric structure and semi-flexibility. It is hard to predict dynamical behavior of a newly synthesized polymer, at given thermodynamic conditions, within the reptation formalism. Furthermore, none of the implementations of the reptation model or the model itself can be used to explain experimental data of the dynamics for short, unentangled systems. This is a consequence of the phenomenological character of the reptation model, which is not general built on a first-principles formalism, and so it cannot be easily implemented to treat more complex systems.
It seems clear that the development of microscopic theories for entangled dynamics could be a necessary step to alleviate this problem. The question is, then, about which kind of microscopic formalism should be adopted. Some considerations could help address this question. 

1) Although the concept of entanglements is quite intuitive, a microscopic theory of entanglements and their effect on polymer dynamics is difficult to formalize because entanglements are essentially a many-body problem. Recent experiments\cite{Granick} and simulations\cite{Larson} have shown that the main effect of entanglements is the confinement of the chain dynamics to a primitive path, in agreement with the original model by Doi and Edwards.\cite{doiedw} From the general point of view, this finding supports the traditional picture of a polymer confined by the presence of entanglements into moving inside a "tube" and explains the success of the reptation model. However, the experimental studies also show that the physics of entanglements is more complex than the simple ``tube" picture of the reptation models. The "tube" is not a static structure, but is dynamic in nature, and a more microscopic approach to entanglements has to be developed.\cite{Ne,Ne1,zcode,Ken}

2) The lack of flexibility of the reptation model when applied to different experimental conditions (temperature, density, polymer degree of semiflexibility, concentration)  arises from its mean-field character, the single chain is treated explicitly but the surrounding chains are described in an approximated way as a ``confining tube". In reality, entanglements involve local persistent contact \textit{between monomers belonging to different chains} and a more effective representation of the key components in this process should involve the dynamics of at least a pair of polymers. Accounting for entanglements as interacting chains was the motivation behind previous approaches, including the double reptation model by des Cloizeaux\cite{desCloizeaux} and the dynamics of two entangled stiff polymers by Szamel.\cite{grezorg,Ken} By formalizing a theory of the dynamics for entangled polymers with explicit intra- and inter-molecular potentials, the effect of the thermodynamics on the dynamics can be properly accounted for, as we show in this paper.\cite{ClarkPRL,Jay2}

3) Up to now, different equations have to be used to describe the motion of a polymeric liquid with  increasing chain length because the Rouse equation and the ``reptation" model are formally incompatible theories. Interestingly, neither the thermodynamics nor the structural properties of the polymeric liquid show any discontinuity at the transition from unentangled to entangled dynamics, indicating that the microscopic formalism expressed at the level of the forces  should be identical in the two regimes. It is reasonable to argue that with increasing chain length different physical phenomena emerge, which are always present in the polymeric liquid, but whose relative relevance changes going from short to long chains. This observation implies that it should be possible to develop one theoretical formalism that properly represents the change in the balance of the forces acting on the polymer at the transition between the two dynamical regimes and which well represents the dynamics in both regimes.

More specifically, while entanglements are always present, in the case of short chains their number per chain is small and their effect on the dynamics should be negligible because the chains can inter-diffuse faster than they experience the physical constraint of entanglements. In long chains, instead, the entanglements dominate the dynamics because the number of entanglements per chain is larger, chain inter-diffusion becomes increasingly slow with increasing degree of polymerization, and chains are not able to disentangle rapidly enough.

4) From a physical perspective both unentangled and entangled polymer melts are complex fluids at low Reynolds number, i.e. they are overdamped systems.\cite{lifealow} The  microscopic formalism that properly represents their dynamics is a Langevin-type of equation in the coordinates of the monomeric unit or the equivalent description of the Fokker-Plank-Smoluchowski equation.\cite{doiedw} A Langevin equation that properly represents the balance of the different inter- and intra-molecular forces playing a role in the polymer melt should provide a good description of the dynamics of the system and good agreement with the different types of experimental data, in the full range of the variable degree of polymerization. 

Other Generalized Langevin Equations were developed in the past to describe entangled dynamics, where the slowing down due to entanglements was formulated through the perturbative solution of the memory function.\cite{Ronca,Hess, Schweizer, Schweizer0,Schweizer1, Fatkullin,Fatkullin1} The approach we follow goes towards a different direction based on the consideration that, in the cases in which it is possible to include a proper description of all the relevant forces that act on the system in the linear part of the Langevin equation, this procedure de facto minimizes the correction due to the memory function and the latter can be conveniently discarded. Because the solution of the memory function requires the simplification of four-point distribution functions into a combination of two-point distribution functions, which is necessarily approximated, having the relevant physics in the linear part of the equation is an advantage. Furthermore, considering the changes in the values of the scaling exponents of diffusion and viscosity as a function of chain length, due to the presence of entanglements, the choice of accounting for the effect of entanglements by using a simple perturbative approach to the unentangled dynamics seems questionable.

The microscopic molecular approach presented here describes the time evolution of the space coordinates of a group of interacting chains, which at initial time are interpenetrating, as they interdiffuse in the field of the surrounding molecules. Selecting a group of slow-moving, interacting chains is justified by the fact that polymers are glass formers and their dynamics present interchanging regions of slow and fast moving molecules. By taking advantage of the observed separation of timescales between slow moving and fast moving regions, we  apply projection operator techniques and derive the equations of motion for the subensemble of slow interacting macromolecules, moving in the field of the fast ones.\cite{marina} In the resulting Langevin Equation, intramolecular and intermolecular  forces guide the dynamical behavior of the polymers as chains interdiffuse and finally become dynamically uncorrelated. 

The theory is microscopic and system specific because it describes the dynamics of polymers with specific flexibility \textit{at the given experimental thermodynamic conditions},\cite{Clark} in this way it situates the dynamics of the polymer melt in the given point of its phase diagram. 
At the intramolecular level, each polymer is described  as a finite chain of monomers connected by springs where the local rigidity due to conformational barriers is accounted for by using a constraint in the angle between bonds, i.e. the freely rotating chain model. This model is formally equivalent to the worm-like chain model, which has been extensively used to describe stiff molecules such as DNA.\cite{DNA} Other models of local barrier crossing and semi-flexibility could be easily implemented,\cite{jeremyC,flory, blumen} but for the purpose of this work the proposed model is quite accurate.  The entanglements are represented by a confining potential, which limits the relative separation distance between two monomers initially in contact, until the chain diffuses outside the entanglement constraint. The formalism is identical in both the unentangled and the entangled regimes, but the confinement due to entanglements become relevant only for those samples for which the chains are long enough for their diffusion to be slower than the time needed for the chains to disentangle.

We selected polyethylene melts (also reported in the literature as PEB-2 samples) to test this theory because data of the dynamics from Neutron Spin Echo experiments are available for an extended range of the degree of polymerization, covering both unentangled and entangled dynamics.\cite{richter1,Zamponi}  This enables us to complete a comprehensive study of the dynamics across the transition from unentangled to entangled dynamics and well into the two opposite regimes. The comparison of the theory with scattering data of unentangled dynamics was reported in an earlier publication,\cite{Zamponi} where the theory was used to test the theoretical description of the cooperative dynamics, while including effects of chain semiflexibility, but not yet the confining potential due to entanglements. A brief overview of the approach presented here has been published in a short publication.\cite{marPRE}

In this paper the complete theory with the entanglement-generated confining interaction between chains is formalized and compared with the full set of data, both in the unentangled and in the entangled regimes. The presence of entanglements does not modify the theoretical predictions for unentangled samples, where $N < N_e$, but provides a better agreement with the data in the crossover region, where our previous work using the theory of cooperative dynamics for unentangled chains showed some level of disagreement.\cite{Zamponi} 

\section{Dynamics of a group of $n$ interacting chains in a melt}

The formalism presented in this paper builds on the theory for the Cooperative Dynamics of polymeric liquids\cite{marina,marina1,marina2} which describes how a group of chains, initially in an interpenetrating configuration, inter-diffuse while their motion becomes progressively less correlated and finally recovers single-chain Fickian diffusion. The presence of intermolecular interactions couples the interchain dynamics until the molecules interdiffuse a distance larger than the range of the interaction potential. 
This approach is general and applies to polymer melts of different local flexibility and monomeric chemical nature as well as polymers of increasing degree of polymerization, including both unentangled and entangled dynamics. The theory for unentangled chains was presented in a series of early papers\cite{marina,marina1,marina2} and its extension to include entangled polymer dynamics was briefly illustrated in a recent short paper\cite{marPRE}. Here, we provide a detailed description of the formalism and an analysis of the different contributions to the experimental chain dynamics as they emerge from comparison with the theory.

Given a group of $n$ polymers initially occupying a volume of the order of the polymer radius-of-gyration, $R_g$, the time evolution of a generic monomer $a$ in the polymer $j$ is governed by a Langevin equation in the space coordinates 
\begin{eqnarray}
\zeta_{eff} \frac{d \mathbf{r}_a^{(j)}(t)}{d t} & = &- \frac{1}{ \beta} 
\frac{\partial}{\partial \mathbf{r}_a^{(j)}(t)} \ln\left[ \prod_{i=1}^n \Psi \left(\{ \mathbf{r}_a^{(i)}(t) \}  \right)   \prod_{k<i=1}^n g \left(\{ \mathbf{r}_a^{(k)}(t), \mathbf{r}_a^{(i)}(t) \}  \right)   \right] + \mathbf{F}_a^{(j)}(t) \ ,
\label{piripo} 
\end{eqnarray}
where $\beta^{-1}=k_BT$ with $k_B$ the Boltzmann constant and $T$ the temperature in Kelvin. Also, $\zeta_{eff}$ is the effective monomer friction coefficient, $\Psi \left(\{ \mathbf{r}_a^{(i)}(t) \}  \right)$ is the intramolecular distribution function, and $g \left(\{ \mathbf{r}_a^{(k)}(t), \mathbf{r}_a^{(i)}(t) \}  \right)$ is the intermolecular distribution function  for the $2N$ monomers in a pair of polymers ($k$ and $i$) of degree of polymerization $N$. 
$\mathbf{F}_a^{(j)}(t)$ is the random force, due to collisions with the $n$ "tagged" molecules, of the ``other" macromolecules that do not belong to the sub-ensemble of the  dynamically correlated ones. As the motion of the correlated chains evolves in time, inter-diffusion progressively decouples their dynamics until the motion of each chain becomes fully independent and stochastic. This happens when each pair of macromolecules moves a relative distance larger than the range of the potential, i.e larger than the range of the correlation-hole of the pair distribution function, $g \left(\{ \mathbf{r}_a^{(k)}(t), \mathbf{r}_a^{(i)}(t) \}  \right) $, which is of the order of the size of the molecule, $R_g$. For distances much larger than $R_g$, the pair distribution function $g \left(\{ \mathbf{r}_a^{(k)}(t), \mathbf{r}_a^{(i)}(t) \}  \right) = 1$, and Eq.(\ref{piripo}) recovers a Rouse-like equation of motion for a group of $n$ chains that are moving independently.

Eq.(\ref{piripo}) is derived\cite{marina} under the assumption that the polymer melt is a dynamically heterogeneous liquid with interconverting regions of slow and fast dynamics. Taking advantage of the separation of time scales between the two dynamics, Eq.(\ref{piripo}) is formally derived by projecting the Liouville equation for the whole liquid onto the coordinates of the sub-ensemble of $n$ slow-moving molecules.\cite{marina}
In Eq.(\ref{piripo}) the many body intermolecular distribution function is replaced with the product of pair distribution functions, in the spirit of a Kirkwood-like superposition approximation. The resulting Langevin equation is formally analogous to the Rouse equation, with the  difference being that pair intermolecular forces resulting from the projection operator formalism,  $K[r(t)] \mathbf{r}(t)$, are explicitly accounted for in the ``linear" part of the equation. For the monomer $a$ belonging to polymer $j$, which is interacting with polymer $i$, the equation of motion
\begin{eqnarray}
\label{eq:langevin}
\zeta_{eff} \frac{d \mathbf{r}_a^{(j)}(t)}{d t} & = & \frac{3}{ \beta l^2} 
\frac{\partial^2  \mathbf{r}_a^{(j)}(t)}{\partial a^2}-(n-1)K[r(t)] \mathbf{r}_a^{(j)}(t)
+\sum_{i \neq j}^{n} K[r(t)] \mathbf{r}_{cm}^{(i)}(t) 
\label{eq:langevinnomemory} 
+ \mathbf{F}_a^{Q(j)}(t)  \ .
\end{eqnarray}

For semiflexible chains of finite length the intramolecular potential is  a function of the structural matrix $\mathbf{A}$, with a Gaussian distribution of intramolecular site positions
\begin{eqnarray}
\Psi \left( \mathbf{r}(t)  \right)  =\left( \frac{3 }{2 \pi l^2}    \right)^{3/2} 
exp \left( \frac{3 k_B T}{2} \frac{ \mathbf{r}(t) ^T \mathbf{A}_{intra} \mathbf{r}(t)  }{l^2} \right) \ ,
\label{distr}
\end{eqnarray} 
where the bond length is $l=|\mathbf{l}|$.

The structural matrix $\mathbf{A}_{intra}$ contains information on how different monomers are connected to form the chain.\cite{doiedw} For semiflexible chains the matrix represents the statistical correlation between bonds and can be calculated following different models,\cite{esterpaper} for example here we use a simple freely-rotating-chain model.\cite{marina} The radius-of-gyration is  $R_g^2=Nl_{eff}^2/6$, with the effective segment length  $\textbf{l}_{eff} \approx \textbf{l} \ (1 + g)/(1 - g)$ and $g=- <\cos \theta>$. The parameter $\theta$ is the statistical angle between two consecutive bonds and accounts for the local semi-flexibility of the polymer.\cite{perslength} The formalism holds for polymeric chains with a variable monomeric structure with the theta parameter depending on the type of polymer under study.\cite{FloryRIS,Ivan,PRELiub,JCPLiub}

Eq.(\ref{eq:langevinnomemory}) can be written in a simple form as\cite{marina,marina1,marina2}
\begin{eqnarray}
\zeta_{eff} \frac{d \mathbf{r}(t)}{dt}= - k_s \mathbf{A}_{tot} \mathbf{r}(t) + \mathbf{F}^{Q}(t) \ ,
\end{eqnarray}
where $\mathbf{A}_{tot} $ is a matrix of dimension $nN \times nN$
\begin{eqnarray}
\mathbf{A}_{tot}=\mathbf{A}_{intra}+\frac{K[r(t)] }{k_s} \mathbf{A}_{inter} \ ,
\end{eqnarray}
and $k_s=3 k_BT/l^2$ is the intramolecular ``spring" constant, while $K[r(t)] [\mathbf{r}_a^{(j)}(t) -\mathbf{r}_{cm}^{(i)}(t)]$ is the time-dependent intermolecular force, which is a function of the relative distance between pairs of interpenetrating macromolecules. Its form will be specified in Section \ref{potential}.

In each chain the dynamics of the center-of-mass can be formally separated from the internal dynamics when the equation of motion of the monomers is formulated relatively to the center-of-mass of the molecule. This is equivalent to setting the center-of-mass of the molecule to be the center of reference for the system of chosen coordinates. In normal modes, which are linearly independent, translation is represented by the zero mode.  In the framework of our approach, the center-of-mass dynamics, and the zero mode of motion, separate into relative and collective contributions.\cite{marina}

The intramolecular monomer distribution accounts for local semiflexibility and chain connectivity however, in this Rouse-like dynamical model, polymers are still represented as phantom chains and monomers belonging to the same molecule can cross each-other. This apparent neglect of excluded volume interactions is based on Flory's hypothesis that the chain statistics in a melt is unperturbed because intra and intermolecular excluded volume interactions tend to compensate each other.

In a recent series of papers we have shown how a liquid of phantom Rouse chains interacting through the effective soft sphere potential correctly follows the equation of state of the polymer melt and reproduces pressure, compressibility, and free energy differences as measured for the polymer liquid described, for example, by MD simulations with atomistic resolution.\cite{Jay2}

Another important approximation in our model is the assumption that chain self-entangling can be discarded as a higher order effect. It is know that in reality long chains do self-entangle and that this phenomenon is important, for example, in the dynamics of translocation of DNA.\cite{DNAknots} However, in a liquid of polymers self-entangling gives only higher order corrections to the simple intermolecular entangled dynamics and can be in first approximation discarded. We see that our approximated description is  a good representation of entangled and also unentangled chain dynamics. This indicates that the phenomenon of self-entangling is not the dominant contribution to the slowing down of the entangled dynamics, at least for high polymer concentrations.

The intramolecular matrix $\textbf{A}_{intra}$ is a block diagonal matrix with $n$ equivalent block matrices $\textbf{A}$ of dimension $N \times N$. This matrix $\textbf{A}$ is the matrix of the intramolecular structure and interaction and it reduces to the Rouse matrix for totally flexible chains.\cite{doiedw,marina} The intermolecular matrix is instead formed by $n$ equivalent block matrices on the diagonal $\textbf{A}_S=(n-1)\textbf{1}$, with $\textbf{1}$ the identity matrix, and $n(n-1)$ identical matrices off the diagonal, whose elements are $(\textbf{A}_U)_{i,j}=-1/N$ for any $i$ and $j$. The intermolecular matrix represents the effective interactions between the center-of-mass of each macromolecule involved (the diagonal terms)  and the monomers belonging to the other interpenetrating chains (the off-diagonal elements).  While the interactions are harmonic, the related spring constant is time dependent and, at each time interval, is optimized through a self-consistent procedure until convergence is achieved. The resulting intermolecular spring constant changes in magnitude with time, representing the time evolution of the intermolecular correlation among interpenetrating chains as the molecules inter-diffuse. 

Taking advantage of the fact that, statistically, each chain in the liquid is equivalent and that the liquid is isotropic, the solution of the set of coupled equations reduces to the study of the dynamics of a pair of chains. More specifically, it reduces to the study of the equations of motion for both the relative dynamics of a generic pair of macromolecules and the collective dynamics of their center of mass. Formally this corresponds to the diagonalization of the block structure of the ``super-matrix" $\textbf{A}_{tot}$ by introducing the similarity transformation matrix, $\mathbf{T}$, which acts on the blocks as
\begin{eqnarray}
\mathbf{T} = \left[  \begin{array} {cccccccc} \beta_1 & \beta_2 & ...  &... & \beta_{n-1} & 
1/\sqrt{n}  \\
- \beta_1 & \beta_2 & ... & ... & \beta_{n-1} & 1/\sqrt{n}   \\
... & ... & ... & ... & ... & ...  \\
0 & 0 & 0 & ... & - (n -1) \beta_{n-1}  & 1/\sqrt{n}   \\
\end{array}  \right ]  \otimes \mathbf{1}\ ,
\end{eqnarray}
with $\beta_i=[i(i+1)]^{-1/2}$ and $\mathbf{1}$ is the $N \times N$ identity matrix.
The matrix $\mathbf{T}$ obeys the condition that $\mathbf{T}^T \mathbf{T}=\mathbf{I}$. 

By introducing the matrix  $\mathbf{T}$, the generalized Langevin equation reduces to $n-1$ identical equations in the relative monomer coordinates, $\mathbf{r}_D(t)$, and one equation in the collective monomer coordinates, $\mathbf{r}_N(t)$,
\begin{eqnarray}
\zeta_{eff} \frac{d \mathbf{r}_D(t)}{dt}= \mathbf{A}_D \mathbf{r}_D(t) + \mathbf{F}_D^{Q}(t) \  \\
\zeta_{eff} \frac{d \mathbf{r}_N(t)}{dt}= \mathbf{A}_N \mathbf{r}_N(t) + \mathbf{F}_N^{Q}(t) \ .
\end{eqnarray}
As the total matrix $\textbf{A}_{tot}$ becomes block diagonal with $n-1$ identical blocks, $\mathbf{A}_D$, and one final block given by the matrix $\mathbf{A}_N$,
\begin{eqnarray}
\mathbf{A}_D=\frac{\beta l^2 K[r(t)]}{3} [ (n-1)\mathbf{1}+\mathbf{Q}_0 \mathbf{Q}_0^T  ] + \mathbf{A} \ ,
\label{AD}
\end{eqnarray}
and

\begin{eqnarray}
\mathbf{A}_N=\frac{(n-1)\beta l^2 K[r(t)]}{3} [ \mathbf{1}-\mathbf{Q}_0\mathbf{Q}_0^T ] + \mathbf{A} \ ,
\end{eqnarray}
where the eigenvector $\textbf{Q}_0$ is the first eigenvector of the Rouse matrix, defined as $\textbf{Q}_0^T=N^{-1/2}(1, 1, ...., 1)$. Adopting an identical friction coefficient, $\zeta_{eff}$, for both relative and collective coordinates implies that the memory function in the original generalized Langevin Equation can be neglected,\cite{marina,marina1,marina2} this is a valid approximation when the most relevant dynamics is already properly accounted for in the linearized part of the equation. We argue that this is the case in this approach because all the relevant interactions are explicitly represented in the equation, while the good agreement that this approach displays when compared with experiments further supports this conclusion.

The transformed coordinates have $n-1$ degenerate vectors describing the relative polymer positions, $ \mathbf{r}_D(t)$, and one
vector, $ \mathbf{r}_N(t)$, defining the collective center-of-mass coordinate,
\begin{eqnarray}
 \left[ \begin{array}{c} \mathbf{r}_D(t) \\ \mathbf{r}_D(t) \\  .... \\ 
 \mathbf{r}_D(t) \\ 
 \mathbf{r}_N(t) \\ \\ 
 \end{array}  \right ] 
  =\mathbf{T}^T\mathbf{r}(t)= \left[ \begin{array}{c}
2^{(-1/2)} \left( \mathbf{r}^{(1)}(t)- \mathbf{r}^{(2)}(t) \right) \\ 
6^{(-1/2)} \left( \mathbf{r}^{(1)}(t) + \mathbf{r}^{(2)}(t) - 2 \mathbf{r}^{(3)}(t)\right) \\
 ...  \\  
\left[ n(n-1)\right]^{-1/2}  \left( \sum_{i=1}^{n-1} \mathbf{r}^{(i)}(t)  - (n-1) \mathbf{r}^{n}(t) \right) \\
(n)^{-1/2} 
\left[ \sum_{i=1}^{n} \mathbf{r}^{(i)} \right]
\end{array}  \right ] \ .
\end{eqnarray}
The problem of solving the set of $n$ coupled equations of motion, each containing $N$ coupled intramolecular equations, reduces by this procedure to the solution of two $N \times N$ equations of motion for the relative and collective dynamics of $n$ chains. Here, the relative and collective eigenvectors are related to the full matrix eigenvectors by the same similarity transformation, $\textbf{T}$, and are just linear combinations of the eigenvectors for the single chain dynamics. In practice, the solution of the correlated dynamics requires only the diagonalization of the single chain intramolecular matrix because of the chosen representation for the effective intermolecular interactions, this reduces greatly the computational cost of these calculations.   
   
In the next step, the dynamics of the center-of-mass of each chain is decoupled from the internal dynamics through Fourier transform into the conventional normal mode description, where
$\mathbf{r}_a^{(j)}(t)= \sum_p \mathbf{Q'}_{a,p}^{(j)} \mathbf{x}_{p}^{(j)}$. The relative mode coordinates are defined as $\mathbf{\xi}_{p}(t)=\sum_a \left[ \mathbf{Q}_{a,p} \right]^{-1} \mathbf{r}_{D,a}(t)$, while the collective mode coordinates are $\mathbf{\chi}_{p}(t)=\sum_a \left[ \mathbf{Q}_{a,p} \right]^{-1} \mathbf{r}_{N,a}(t)$.     
In relative and collective normal mode description, the Langevin equation for the internal dynamics ($p=1,\ 2, ...,N-1$) reduces to 

\begin{eqnarray}
\label{eq:normalm1}
\zeta_{eff} \frac{d \mathbf{\xi}_{p}^{(j)}(t)}{d t} & = & -  \left( \frac{3}{ \beta l^2} \lambda_p+(n-1)K[r(t)]  \right) \mathbf{\xi}_{p}^{(j)}(t)
- K[r(t)] \mathbf{\xi}_{0}^{(i)}(t)  + \mathbf{F}_{\xi,p}^{Q(j)}(t)  \ , \\ 
\zeta_{eff} \frac{d \mathbf{\chi}_{p}^{(j)}(t)}{d t} & = & -  \left( \frac{3}{ \beta l^2} \lambda_p+(n-1)K[r(t)] \right) \mathbf{\chi}_{p}^{(j)}(t)
+\sum_{i \neq j}^{n} K[r(t)] \mathbf{\chi}_{0}^{(i)}(t)  + \mathbf{F}_{\chi,p}^{Q(j)}(t) \nonumber \ , 
\end{eqnarray}
where ${\lambda_p}$ are the eigenvalues of the single-chain intramolecular matrix. Approximating the projected force into the unprojected one,\cite{marina} which is a valid approximation when there is clear separation of dynamical timescales between projected variables and unprojected ones, andthe internal chain dynamics is given by a set of uncoupled equations of motion in the mode coordinates
\begin{eqnarray}
\label{eq:normalm11}
\zeta_{eff} \frac{d \mathbf{\xi}_{p}^{(j)}(t)}{d t} & = & - \left( \frac{3}{ \beta l^2} \lambda_{p}+(n-1)K[r(t)]  \right) \mathbf{\xi}_{p}^{(j)}(t)
+ \mathbf{F}_{p}^{\xi}(t) \ , 
\label{eq:normalm11} \\
\zeta_{eff} \frac{d \mathbf{\chi}_{p}^{(j)}(t)}{d t} & = & - \left( \frac{3}{ \beta l^2} \lambda_{p}+(n-1)K[r(t)]  \right) \mathbf{\chi}_{p}^{(j)}(t)
+ \mathbf{F}_{p}^{\chi}(t) \nonumber \ .
\end{eqnarray}

\noindent The related fluctuation-dissipation theorem reads
\begin{eqnarray}
< \mathbf{F}_{p}(t) \cdot \mathbf{F}_{q}(t')> = 6 k_B T \zeta_{eff} \delta(t,t') \delta(p,q)  \ ,
\end{eqnarray}
for both the relative ($\xi_p$) and the collective ($\chi_p$) dynamics.

The solution of Eqs. (\ref{eq:normalm11}) for the internal modes ($p=1, \ 2, ..., N$)  leads to the equations in the relative and collective coordinates, respectively,
\begin{eqnarray}
\label{eq:coordinates}
\mathbf{\xi}_p (t) & = & \mathbf{\xi}_p(0)e^{- t/\tau_{\xi,p}(t)} + e^{- t/\tau_{\xi,p}(t)}  \int_{0}^{t} d \tau \zeta_{eff}^{-1} \mathbf{F}^{\xi}_{p}(\tau) e^{t/\tau_{\xi,p}(\tau)} \ , \\
\mathbf{\chi}_p (t) & = & \mathbf{\chi}_p(0)e^{- t/\tau_{\chi,p}(t)} + e^{- t/\tau_{\chi,p}(t)}  \int_{0}^{t} d \tau \zeta_{eff}^{-1} \mathbf{F}^{\chi}_{p}(\tau) e^{t/\tau_{\chi,p}(\tau)} \ , \nonumber
\end{eqnarray}

with 
\begin{eqnarray}
\label{tdsc}
\frac{t}{\tau_{\xi,p}(t)}=\frac{k_s \lambda_{p}t}{\zeta_{eff}}+\frac{(n-1)}{\zeta_{eff}}\int_0^t K[r(t')]dt' \ ,  \\
\frac{t}{\tau_{\chi,p}(t)}=\frac{k_s \lambda_{p}t}{\zeta_{eff}}+\frac{(n-1)}{\zeta_{eff}}\int_0^t K[r(t')]dt' \ , \nonumber 
\end{eqnarray}

\noindent where $\tau_{sc,p}=\zeta_{eff}/(k_s \lambda_{p})$ is the characteristic relaxation time of the single chain without entanglement effects.
The interaction due to entanglements acts at the level of the monomers and  enters the equations of motion of the internal modes as an effective force, $K[r(t')]$. The intermolecular force between monomers due to entanglements is zero until the distance between two monomers that are initially in contact, reaches the entanglement distance $d$. If the chain is fully relaxed before this characteristic time at which chains start to feel the presence of entanglements, i.e. if $\tau_{Rouse} < \tau_e$ with $\tau_{Rouse}=\tau_{sc,p=1}$ the relaxation time of the longest single-chain mode, and  $\tau_e=d^2/D$ the entanglement time, the entanglements are not affecting the dynamics. This is the case for unentangled chains in our model.

For the center of mass dynamics, where $p=0$ and $\lambda_0=0$, the dynamics in relative and collective coordinates follows the equation of motion
\begin{eqnarray}
\label{eq:normalm12}
\zeta_{eff} \frac{d \mathbf{\xi}_{0}(t)}{d t} & = & - n K_0[r(t)] \mathbf{\xi}_{0}(t)
+ \mathbf{F}_{0}^{\xi}(t) \ ,  \\
\zeta_{eff} \frac{d \mathbf{\chi}_{0}(t)}{d t} & = & \mathbf{F}_{0}^{\chi}(t) \nonumber \ , 
\end{eqnarray}
with the fluctuation-dissipation theorem defined as
\begin{eqnarray}
\label{perrocche'}
< \mathbf{F}_{0}^{\xi}(t) \cdot \mathbf{F}_{0}^{\xi}(t')> & = & 6 n(n-1) k_B T \zeta_{eff} \delta(t,t')  \ , \label{lillibet} \\
< \mathbf{F}_{0}^{\chi}(t) \cdot \mathbf{F}_{0}^{\chi}(t')> & = & 6 n  \ k_B T \zeta_{eff} \delta(t,t')  \ . \nonumber
\end{eqnarray}
Eq.(\ref{perrocche'}) emerges as the multibody structural distribution function is approximated by a product of pair distribution functions. Higher-order coupling terms would enter the friction coefficient through the memory function and in our formalism they lead to an effective friction coefficient. The quality of the agreement between the theory and the experiments, and between the theory and the simulations, suggests that this simple approximation, when coupled with the self-consistent procedure, represents well the many-body nature of the interactions and the correlation of the dynamics.

The intermolecular force that acts at the level of the center-of-mass, $K_0[r(t)]$, results from the random propagation of the monomer-monomer interaction between any pair of interpenetrating chains through the liquid.\cite{Clark,Ivan} A liquid of phantom Rouse chains is fully compressible and thermodynamically unrealistic: in our approach by including the intermolecular interaction in the Rouse model, or in its implementations, the liquid of phantom chains recovers the correct compressibility and equation of state and the theory describes the dynamics of a polymer liquid \textit{under controlled thermodynamic conditions}.\cite{Clark,ClarkPRL,Jay2}

By including the center-of-mass potential the relative dynamics is given by
\begin{eqnarray}
\mathbf{\xi}_0 (t) = 
 \mathbf{\xi}_0(0)e^{- t/\tau_{\xi,p}(t)} + e^{- t/\tau_{\xi,p}(t)}  \int_{0}^{t} d \tau \zeta_{eff}^{-1} \mathbf{F}^{\xi}_{p}(\tau) e^{t/\tau_{\xi,p}(\tau)} \ ,
\end{eqnarray}
with 
\begin{eqnarray}
\frac{t}{\tau_{\xi,0}(t)}=\frac{n}{\zeta_{eff}}\int_0^t K_0[r(t)]dt' \ .
\end{eqnarray}

In this case the center-of-mass chain dynamics, for either unentangled or entangled polymers, becomes subdiffusive until the time necessary for the chain to move a distance larger than the range of the interaction, when the chain motion becomes uncorrelated and Brownian (see for example the data in Figure \ref{msd}).

When the intermolecular interaction at the center-of-mass level is zero, $K_0[r(t)]=0$, Eq.(\ref{lillibet}) for $\xi_0$ reduces to the diffusion equation and no anomalous center-of-mass motion emerges from this approach. For short, unentangled chains this limit corresponds to the  trivial motion of $n$ independent, inter-diffusing chains, each following the Rouse equation with internal semiflexibility.


\section{Effective Interchain Potentials}
\label{potential}
The theory discussed in this paper describes the time evolution of a sub-ensemble of $n$, entangled or unentangled, correlated macromolecules. The presence of surrounding molecules affects the dynamics by generating viscosity and friction. It also affects the dynamics non-trivially through the effective intermolecular potentials originating both from the presence of the many-body interactions in the liquid and of the entanglements. 

\subsection{Many-body intermolecular potential between chains}
 
In a liquid of neutral polymers, the effective intermolecular potential between chains reflects how local intermolecular monomer-monomer interactions propagate through the medium leading to the effective pair interactions between the center-of-mass of a pair of chains.\cite{Clark,ClarkPRL} The excluded volume intermolecular interaction between monomers generates an effective potential given by  the projection of these many-body interactions through the liquid onto a pair of effective sites, in this case the center-of-mass on each chain. The potential has a long-range repulsive component and a small attractive part, which is largely entropic in nature, and generates from the multiple liquid configurations.

Given that the theory tracks the dynamics of a relatively small number of molecules, $n=\rho R_g^3$, the effective potential is well-approximated by the potential of mean force, which is the potential between two molecules in the field of the others. At the density and temperature of the samples studied here, the effective potential and the potential of mean force are close in magnitude and shape,\cite{Ivan} 
and the diffusive dynamics is mainly affected by the repulsive part of the potential. The whole potential can be well-approximated by a Gaussian repulsive function, with a time-dependent spring constant for the center-of-mass intermolecular force given by\cite{marina,marina1,marina2}
\begin{eqnarray}
K_0[r(t)]\approx - \frac{171}{32} \sqrt{\frac{3}{\pi}} N \frac{\xi_{\rho}}{R_g^3} k_BT
\Big(1 + \frac{\sqrt{2}\xi_\rho}{Rg} \Big) e^{-\frac{75 r(t)^2}{76R_g^2}} \ ,
\end{eqnarray}
with $r^2(t) =<[\xi_0(t)-\xi_0(0)]^2>/N$ the square intermolecular center-of-mass distance between a pair of molecules. This distance evolves in time as the polymers inter-diffuse and  the  repulsive force is calculated at each given time interval by means of a self-consistent procedure until the optimized intermolecular distance converges.
Because in our model the many-body intermolecular potential acts between the center-of-mass of the two polymers, it affects only the zeroth mode of motion of the polymer, which represents the translational diffusion.

\subsection{Entanglement Potential}
Our model accounts also for the effect of entanglements by applying a potential that is zero at any inter-monomer distance  smaller than a characteristic distance, $d$, at which the entangled loop experiences a force that opposes further stretching.  When two monomers, initially at contact, move a relative distance that is larger than the given average value, $d$, the monomers experience a repulsion that tends to confine their relative dynamics. 
At longer times the constraint due to the entanglement relaxes because the potential evolves as the polymers interdiffuse relative to each other. 

In a reptation-like representation this relaxation dynamics corresponds to the diffusing of the polymer outside the tube and to the different mechanisms that relax the constraint due to entanglements, e.g. constraint release and tube fluctuations. In our many-chain model it is not necessary to hypothesize different mechanisms for the relaxation of the dynamics, but they naturally emerge from the solution of the interchain Langevin equation. At a given time interval the time-dependent intermonomer distance, which is included in the intermolecular potentials, is optimized through a self-consistent procedure. As the time interval increases the distance between the monomers, which were initially in contact and entangled, becomes increasingly larger, until the relative motion of the two monomers becomes fully uncorrelated.

In the volume occupied by a chain, $R_g^3$, the chain is statistically in contact with $n-1$ other chains. Entanglements occur between the ``tagged'' chain and the $n-1$ other chains that are interpenetrating. The number of entanglements that the ``tagged'' chain experiences is given by the total number of monomers in the given volume,  $(n-1)N \approx \rho R_g^3$, divided by the number of monomers in a chain segment between a pair of entanglements, $N_e$. The statistical number of entanglements per chain is then $(n-1)N/N_e\approx \rho R_g^3/N_e$. 

The intermonomer potential due to entanglement is zero until the monomers reach a relative  distance comparable to $d$. At that distance the potential is characterized by a repulsion weighted by the probability of finding two monomers in mutually entangled chains at a given distance, $r$. Any given monomer in a polymer melt has a conditional probability $g(r)$ of finding another monomer belonging to a different chain at some distance $r$. This is multiplied by the probability of finding the first monomer in an entangled segment, which is given by the density of entanglements $\rho/N_e $, weighted by the distribution of monomers inside the chain segment delimited by two entanglements. That statistical distribution is assumed to have a Gaussian shape. If we define $\textbf{R}$ as the distance between monomer $a$ and the position of the center of mass of the entangled segment that includes monomer $b$, then  $r=|\textbf{r}|=|\textbf{R} + \textbf{r}_b|$. 
Thus the potential between a pair of monomers, $a$ and $b$, belonging to different chains inside the correlation hole, due to entanglements, is written as
\begin{eqnarray}
V[r,R,t]=  - k_B T \rho/N_e  \ln [ g(r,t)]    \ \ \ for \  R > d \ ,
\end{eqnarray}
and $V(r,R,t)=0$ for $ R \le  d$
where $R$ is the distance between monomer $a$ and a generic monomer $b$ belonging to a segment entangled with the first chain segment.

The intermolecular force constant between monomer $a$ and a generic monomer inside the entangled segment belonging to another chain, due to entanglements, is defined as
\begin{eqnarray}
\label{forcec}
K[R(t)] & \approx & - k_B T \rho/N_e <(1/|r(t)|) \frac{\partial V[r(t)]}{\partial r}  >   \nonumber \\
& = & k_BT \rho/N_e \int d \textbf{r}_{b}(t) (1/|r(t)|)   \Psi(r_b (t)) \frac{\partial g(r,t)}{\partial r}  \ .
\end{eqnarray}
The potential is calculated as an average over the position distribution of the monomer $b$ in an entangled segment of the second polymer,
\begin{eqnarray}
\Psi(r_b (t))=\Big( \frac{3}{2 \pi d^2} \Big)^{3/2} e^{-\frac{3}{2}\frac{\textbf{r}_b^2(t)}{d^2}} \ .
\end{eqnarray}
Because the intermolecular force experienced by the molecules depends on their reciprocal  distance, the equation of motion is far from equilibrium.\cite{manychains} Eq.\ref{eq:langevin} is solved through a self-consistent procedure at a fixed time interval; fluctuations of the intermolecular distance are allowed, but the average intermonomer distance is at each time step the one that is obtained from the optimization. Because the equation is solved at a fixed time interval, we drop in our notation the time dependence  (see Eq.\ref{forcec}) with the understanding that the intermonomer distance is in itself time dependent. 

An analytical expression for the effective force acting between two monomers that belong to two chains mutually entangled can be derived using the thread model representation of the monomer pair distribution function for a liquid of polymer chains. In the thread model a polymer is described as an infinitely thin and infinitely long chain, while the density of the liquid is kept constant.  In the thread model, the PRISM theory\cite{PRISM} gives for the pair distribution function of the monomers inside the volume spanned by an entangled segment the simple analytical expression
\begin{eqnarray}
g(r)=1+\frac{3}{\pi \rho \sigma^2} \Big[  \frac{e^{-|\textbf{r}|/ \xi_{\rho}}}{|\textbf{r}|}  -  \frac{e^{-|\textbf{r}|/ \xi_{d}}}{|\textbf{r}|}   \Big] \ ,
\end{eqnarray}
which describes the average probability of finding another monomer belonging to another entangled segment at some distance $|\textbf{r}|$
Here, $\xi_{\rho}$ is the local density fluctuation screening length, which is related to the liquid packing fraction and the bulk properties of the systems, such as the liquid compressibility. The second characteristic lengthscale is the entangled segment correlation hole lengthscale, which is related to the size of the entangled segment as $\xi_d \approx d/\sqrt{2}$. 

The equation of the force has two type of terms that are related by a simple transformation
\begin{eqnarray}
\frac{1}{r^2}e^{-r/\xi}= -\frac{\partial }{\partial \xi^{-1}}\frac{e^{-r/\xi}}{r^3} \ ,
\end{eqnarray}
 so that the only integral to solve is of the type
 \begin{eqnarray}
 I=- \beta^{-1} \int d \textbf{r}_{b} \Psi(r_b) \frac{e^{-r/\xi}}{r^3}H[R-d] \ .
\end{eqnarray} 
    
By introducing the Fourier transforms
\begin{eqnarray}
\frac{1}{|\textbf{r}|}= -\frac{4 \pi}{(2 \pi)^3}\int d\textbf{k} \frac{e^{i \textbf{k} \cdot \textbf{r}}}{k^2} \ ,
\end{eqnarray}
and 
\begin{eqnarray}
\frac{e^{-|\textbf{r}|/\xi}}{|\textbf{r}|}= -\frac{4 \pi}{(2 \pi)^3}\int d\textbf{k} \frac{ \xi^2e^{i \textbf{k} \cdot \textbf{r}}}{1+k^2 \xi^2} \ ,
\end{eqnarray}
the integral is rewritten as
 \begin{eqnarray}
I=- \beta^{-1}  \ \frac{1}{8 \pi^6}  \int \frac{d \textbf{k}_1}{k_1^2}  \int \frac{d \textbf{k}_2}{k_2^2}  
\int \frac{d \textbf{k}_3}{\xi^{-2}+ k_3^2}
\int d \textbf{r}_b  e^{i \textbf{k} \cdot \textbf{r}} \ e^{ - \mu  r_{b}^2} H[R-d]\ ,
\end{eqnarray}
where $\textbf{k}=\textbf{k}_1 + \textbf{k}_2 + \textbf{k}_3$ and $\mu=\frac{3}{2 d^2}$ with $d=\sqrt{N_e}l$. 

Because the  exponential term is dominant at $k\approx N_e^{-1}$, the entanglement force constant is well approximated as 
\begin{eqnarray}
K[R(t)] \approx k_B T (\rho/N_e) \xi_{\rho} (3\sqrt{3}/\sqrt{2 \pi}) (\xi_d / \xi_\rho)^{2} (1/d^3) e^{-\mu R^2}
H[R-d] \ ,
\end{eqnarray}
so that
\begin{eqnarray}
K[R(t)] \approx  k_B T (\rho/N_e)  (d /\xi_{\rho}) (1/d^2) e^{ - \frac{3 (R-d)^2}{2 d^2}}  H[R-d]\ .
\end{eqnarray}

The confining potential is then given by $V[R(t)]=0$ if $|\textbf{r}_{a}^i -\textbf{r}_b^j| \le d $ and
\begin{equation}
V[R(t)] \approx 
k_B T \rho/N_e   e^{-3(R(t)-d)^2/2d^2} \ .
\end{equation}
for $| \textbf{r}_{a}^i -\textbf{r}_b^j| > d$.
The force is generalized for more than one entanglement distance to $K[R(t)] \approx  k_B T \rho/N_e [R(t) -d]/[R(t) d^2]   < exp[-(R(t)-d)^2/(d^2) ] + exp[-(R(t)-2d)^2/(d^2)]>$, which enters Eq.(\ref{tdsc}). This represents the contribution to the confining potential that a monomer in an entangled segment feels due to the presence of entanglements in adjacent regions inside the primary chain.

The force just described applies at the monomer level to both unentangled and entangled polymer melts. While it has a relevant impact on the dynamics of long chains, it has no effect on the dynamics of short chains because their motion becomes uncorrelated before they move an intermolecular monomer distance comparable to $d$.

\begin{figure}
\includegraphics[width=3.0in]{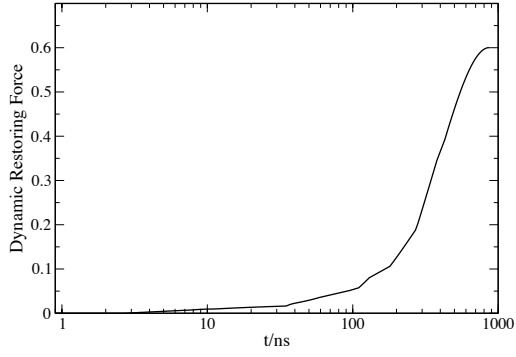} \\
\caption{Effective force that a monomer experiences when it moves a distance comparable to $R(t)$ with respect to another monomer belonging to another chain entangled with the first. The two monomers, which are initially in contact, inter diffuse as a function of time and as they move apart they experience an increasing confining force. This force could be represented by an effective "tube" in a single-chain formalism. The shape of this curve depends on the thermodynamic conditions and on the chemical structure of the polymer.}
\label{figura1}
\end{figure}

Figure \ref{figura1} displays the time-dependent effective force that confines a monomer due to the presence of entanglements.  The force grows smoothly with distance and shows confinement that opposes the free diffusion of the monomer and increases as the monomer approaches the point of entanglement. The range and intensity of the repulsive force is in qualitative agreement with the predictions of simulations and experiments where the confining force was measured for a system where a polymer chain was free to move and the two ends of the chain were not fixed in space.\cite{Larson,Granick} 

In our model, a simple hard-core repulsive interaction limits the maximum relative distance between two chains and leads to an effective force that has the shape of a smooth confining ``tube", in agreement with the common physical picture of ``reptation" and with the shape observed in both experiments and in the analysis of simulations of polymer melts. The resulting smooth shape of the effective interaction is a consequence of the weighting of the hard-core repulsion due to the conformational statistics of the chain and the distribution of the relative distances between monomers. 

\section{Time-correlation functions}
The physical quantities that represent time-dependent properties of a melt of polymers are time correlation functions. Specifically, it is convenient to express the time-correlation-functions in the normal mode coordinates. By applying the eigenvector transformation and the inverse of the similarity transformation matrix, $\mathbf{T}$, the time-correlation-functions give the real-space time-dependent properties of the polymer chains, such as the mean-square displacement of the monomer and of the chain center-of-mass. 
The cross products of the normal mode coordinates, for the \textit{internal modes} where $p=1, 2, ..., N-1$, are given by 
\begin{eqnarray}
<\mathbf{\xi}_p (t) \cdot \mathbf{\xi}_p (0)> & = & <\mathbf{\xi}_p(0)^2> e^{- t/\tau_{\xi,p}(t)} \ , \\
<\mathbf{\chi}_p (t) \cdot  \mathbf{\chi}_p (0)> & = & <\mathbf{\chi}_p(0)^2> e^{- t/\tau_{\chi,p}(t)} \ , \nonumber \\
<\mathbf{\xi}_p (t) \cdot \mathbf{\chi}_p (0)> & = & 0 \nonumber \ ,
\end{eqnarray}

while the self products are
\begin{eqnarray}
<\mathbf{\xi}_p (t)^2> & = & <\mathbf{\xi}_p(0)^2> e^{- 2 t/\tau_{\xi,p}(t)} + \frac{6 k_B T}{\zeta_{eff}} e^{- 2 t/\tau_{\xi,p}(t)}  \int_{0}^{t} d \tau  e^{2 \tau/\tau_{\xi,p}(\tau)} \ , \\
<\mathbf{\chi}_p (t)^2> & = & <\mathbf{\chi}_p(0)^2> e^{- 2 t/\tau_{\chi,p}(t)} + \frac{6 k_B T}{\zeta_{eff}} e^{- 2 t/\tau_{\chi,p}(t)}  \int_{0}^{t} d \tau e^{2 \tau/\tau_{\chi,p}(\tau)} \ . \nonumber
\end{eqnarray}

For a given mode $p$, with  $p=1, 2, ..., N-1$, the relative displacement of two monomers inside a pair of polymers during a time interval $\Delta t=t-t_0$, with $t_0=0$, is given by
\begin{eqnarray}
<[\mathbf{\xi}_p (t)-\mathbf{\xi}_p (0)]^2> & = & <\mathbf{\xi}_p(0)^2> [e^{- t/\tau_{\xi,p}(t)} -1 ]^2 + \frac{6 k_B T}{\zeta_{eff}} e^{- 2 t/\tau_{\xi,p}(t)}  \int_{0}^{t} d \tau  e^{2 \tau/\tau_{\xi,p}(\tau)} \ ,
\end{eqnarray}
while the displacement of the collective normal mode coordinates is
\begin{eqnarray}
<[\mathbf{\chi}_p (t)- \mathbf{\chi}_p (0)]^2> & = & <\mathbf{\chi}_p(0)^2> [e^{- t/\tau_{\chi,p}(t)} -1]^2 + \frac{6 k_B T}{\zeta_{eff}} e^{- 2 t/\tau_{\chi,p}(t)}  \int_{0}^{t} d \tau e^{2 \tau/\tau_{\chi,p}(\tau)} \ .
\end{eqnarray}

If there are not interactions between chains, or between monomers belonging to different chains, the single-chain limit, which is conventionally represented by the Rouse model, can be recovered with $n=1$. The equations simplify as
\begin{eqnarray}
\frac{t}{\tau_{\xi,p}(t)}=\frac{t}{\tau_{\chi,p}(t)}=\frac{k_s \lambda_{p}t}{\zeta_{eff}} =\frac{3 k_B T \lambda_{p}t}{l^2 \zeta_{eff}} \ ,
\end{eqnarray}

\begin{eqnarray}
<\mathbf{\xi}_p (t)^2> & = & <\mathbf{\xi}_p(0)^2> e^{- 2 t/\tau_{\xi,p}(t)} +  \frac{l^2}{\lambda_p} \left(1- e^{- 2 t/\tau_{\xi,p}(t)} \right) \ ,
\end{eqnarray}

\begin{eqnarray}
<[\mathbf{\xi}_p (t)-\mathbf{\xi}_p (0)]^2> & = & <\mathbf{\xi}_p(0)^2> [e^{- t/\tau_{\xi,p}(t)} -1 ]^2 + \frac{l^2}{\lambda_p} \left(1- e^{- 2 t/\tau_{\xi,p}(t)} \right)\ ,
\end{eqnarray}
and given that 
\begin{eqnarray}
<\mathbf{\xi}_p(0)^2> =\frac{l^2}{\lambda_p} \ ,
\end{eqnarray}
we have
\begin{eqnarray}
<\mathbf{\xi}_p(t)^2> =<\mathbf{\xi}_p(0)^2> =\frac{l^2}{\lambda_p} \ ,
\end{eqnarray}
\begin{eqnarray}
<[\mathbf{\xi}_p (t)-\mathbf{\xi}_p (0)]^2> & = & \frac{2l^2}{\lambda_p}[1-e^{- t/\tau_{\xi,p}(t)}  ] = 2 \frac{l^2}{\lambda_p}[1-e^{- 3 k_B T \lambda_p t/(l^2 \zeta_{eff})}  ]  \ ,
\end{eqnarray}
which are the equations in the normal mode coordinates of the Rouse theory. The  intramolecular parameters characteristic of the sample enter the dynamics through the eigenvalues and the friction coefficients. 

For the center-of-mass diffusion, for which $p=0$, the normal mode time-correlation-functions read
  
\begin{eqnarray}
\label{eq:coordinates}
\mathbf{\xi}_0 (t) & = & \mathbf{\xi}_0(0)e^{- n \int_0^tK_0[r(t')]dt' /\zeta_{eff}} + e^{-n \int_0^t K_0[r(t')] dt' /\zeta_{eff}}  \int_{0}^{t} d \tau \zeta_{eff}^{-1} \mathbf{F}_{\xi,p}(\tau) e^{ n \int_0^{\tau} K_0[r(t')] dt' /\zeta_{eff}} \ ,
\end{eqnarray}
and
\begin{eqnarray}
\mathbf{\chi}_0 (t) & = & \mathbf{\chi}_0(0) +   \int_{0}^{t} d \tau \zeta_{eff}^{-1} \mathbf{F}_{\chi,0}(\tau)  \ .
\end{eqnarray}

This leads to
\begin{eqnarray}
<\mathbf{\xi}_0 (t)^2> & = & <\mathbf{\xi}_0(0)^2> e^{-  2t/\tau_{\xi,0}(t)} + \frac{6 n(n-1) k_B T}{\zeta_{eff}} e^{- 2 t/\tau_{\xi,0}(t)}  \int_{0}^{t} d \tau  e^{2 \tau/\tau_{\xi,0}(\tau)} \ , \\
<\mathbf{\chi}_0 (t)^2> & = & <\mathbf{\chi}_0 (0)^2>+  \frac{6 n k_B T}{\zeta_{eff}} t \ , \nonumber
\end{eqnarray}
and for the displacement in relative and collective coordinates
\begin{eqnarray}
<[\mathbf{\xi}_0 (t)-\mathbf{\xi}_0 (0)]^2> & = & <\mathbf{\xi}_0(0)^2> [e^{-  t/\tau_{\xi,0}(t)}-1]^2 + \frac{6 n(n-1) k_B T}{\zeta_{eff}} e^{- 2 t/\tau_{\xi,0}(t)}  \int_{0}^{t} d \tau  e^{2 \tau/\tau_{\xi,0}(\tau)} \ , \\
<[\mathbf{\chi}_0 (t)-\mathbf{\chi}_0 (0)]^2> & = & \frac{6 n k_B T}{\zeta_{eff}} t \ , \nonumber
\end{eqnarray}
where the correlation functions at initial time are  $<\mathbf{\xi}_0 (0)^2> =6 n (n-1) k_BT/\zeta_{eff}$ and $<\mathbf{\chi}_0 (0)^2> =6 n k_BT/\zeta_{eff}$.

In the limit of non interacting polymers, the relative normal mode description is
\begin{eqnarray}
\mathbf{\xi}_0 (t) & = & \mathbf{\xi}_0(0) +   \int_{0}^{t} d \tau \zeta_{eff}^{-1} \mathbf{F}_{\xi,0}(\tau)  \ ,
\end{eqnarray}
which gives
\begin{eqnarray}
<\mathbf{\xi}_0 (t)^2> & = & <\mathbf{\xi}_0 (0)^2>+  n(n-1) \frac{6 k_B T}{\zeta_{eff}} t \ ,
\end{eqnarray}
while the collective time-correlation-function reduces to the diffusive dynamics
\begin{eqnarray}
<\mathbf{\chi}_0 (t)^2> & = & <\mathbf{\chi}_0 (0)^2>+  n \frac{6 k_B T}{\zeta_{eff}} t \ .
\end{eqnarray}

In \textit{real space} coordinates the  dynamics is defined as a function of the normal modes through the transformation
\begin{eqnarray}
\mathbf{r}^i_a(t) & = & \sum_{p=0}^{n(N-1)}  \mathbf{Q'}_{a,p} \mathbf{x}^i _p(t)= \sum_{k=1}^n \frac{1}{\sqrt{k(k+1)}}  \sum_{p=0}^{N-1}  (\mathbf{Q}_\xi)_{a,p} [\mathbf{\xi}_p(t)] + \frac{1}{\sqrt{n}}\sum_{p=0}^{N-1}  (\mathbf{Q}_\chi)_{a,p} [\mathbf{\chi}_p(t)] \ .
\end{eqnarray}
The eigenvectors are orthonormal, with $(\mathbf{Q}_\xi)=(\mathbf{Q}_\chi)=\mathbf{Q} $, and the  condition applies that
$ \mathbf{Q}^T  \mathbf{Q}= \mathbf{1}$.
Using the identity $\sum_{k=0}^{n-1} [k (k+1)]^{-1}=(n-1)/n$ the time correlation functions simplify as follow

\begin{eqnarray}
&& <[\mathbf{r}^i_a(t) -\mathbf{r}^i_a(0)]^2>  =  \frac{n-1}{n}  \sum_{p=0}^{N-1} \mathbf{Q}^2_{a,p} <[\mathbf{\xi}_p(t)-\mathbf{\xi}_p(0)]^2> +  \frac{1}{n}  \sum_{p=0}^{N-1} \mathbf{Q}^2_{a,p} <[\mathbf{\chi}_p(t)-\mathbf{\chi}_p(0)]^2>  \ , \\
&& <[\mathbf{r}^i_a(t) -\mathbf{r}^i_a(0)]\cdot [\mathbf{r}^j_b(t) -\mathbf{r}^j_b(0)]>  =  -\frac{1}{n}  \sum_{p=0}^{N-1} \mathbf{Q}^2_{a,p} <[\mathbf{\xi}_p(t)-\mathbf{\xi}_p(0)]^2> +  \frac{1}{n}  \sum_{p=0}^{N-1} \mathbf{Q}^2_{a,p} <[\mathbf{\chi}_p(t)-\mathbf{\chi}_p(0)]^2> \ . \nonumber
\end{eqnarray}

The increment in the distance between two monomers belonging to different chains during a time interval $\Delta t= t- t_0$, with $t_0=0$, is given by
\begin{eqnarray}
<[(\mathbf{r}^i_a(t) -\mathbf{r}^i_a(0)) & - & (\mathbf{r}^j_b(t) -\mathbf{r}^j_b(0))]^2>  = \\ 
& +  & \frac{1}{n} \sum_{p=0}^N [(n-1)Q^2_{a,p}+(n-1)Q^2_{b,p}+2 Q_{a,p}Q_{b,p}] <[\mathbf{\xi}_p(t)-\mathbf{\xi}_p(0)]^2>   \nonumber \\ & + & \frac{1}{n} \sum_{p=0}^N [Q_{a,p}-Q_{b,p}]^2  <[\mathbf{\chi}_p(t)-\mathbf{\chi}_p(0)]^2> \ . \nonumber
\end{eqnarray}

The approximate expression for the distance between two monomers, $a$ and $b$,  belonging to two different chains is
\begin{eqnarray}
<(\mathbf{r}^j_b(t)-\mathbf{r}^i_a(t))^2> \approx <(\mathbf{r}^j_b(0)-\mathbf{r}^i_a(0))^2>  + <[(\mathbf{r}^i_a(t) -\mathbf{r}^i_a(0)) & - & (\mathbf{r}^j_b(t) -\mathbf{r}^j_b(0))]^2>  
\end{eqnarray}
as 
\begin{eqnarray}
r(t')-r_0 & \approx & <[(\mathbf{r}^i_a(t) -\mathbf{r}^i_a(0)) -  (\mathbf{r}^j_b(t) -\mathbf{r}^j_b(0))]^2> ^{1/2} \ .
\end{eqnarray}

From the equations for the zero normal modes, the time evolution of the position of the center-of-mass for a single chain inside the ensemble of initially correlated $n$ molecules is given by the linear combination of relative and collective contributions as
\begin{eqnarray}
\Delta R^2(t)=<[\mathbf{R}_{cm}(t)-\mathbf{R}_{cm}(0)]^2> & = & n^{-2}N^{-1} [<[\mathbf{\xi}_0 (t)-\mathbf{\xi}_0 (0)]^2>  + <[\mathbf{\chi}_0 (t)-\mathbf{\chi}_0 (t)]^2>] \ ,
\end{eqnarray}
which in the limit of non-interacting chains correctly reduces to the free diffusion of the single chain, i.e. the center-of-mass motion in the Rouse model,  
\begin{eqnarray}
\Delta R^2(t)= \frac{6 k_B T}{\zeta_{eff}} t \ .
\end{eqnarray}

\section{Solving the Langevin Equation}
The effective intermolecular potentials that enter the set of $n$ coupled Langevin equations are functions of the distance between chains. The entanglement potential acts between monomers belonging to two different polymers in the sub-ensemble of chains that are interpenetrating at initial time. Because the intermonomer distance change with time as the polymers inter-diffuse, the related potential is also time dependent. 

The Langevin Equation (LE) is solved  first for the zero mode, which describes center-of-mass diffusion. From the solution of the LE the statistical distance between the center of mass of a pair of polymers is calculated. This distance enters the intermolecular potential in the relative and collective dynamics of the $p=0$ mode. From the solution of the $p=0$ equations the center-of-mass average distance is calculated and optimized through convergence of the self-consistent procedure. Once the inter-polymer distance is optimized, the average interchain monomer-monomer distance is also optimized.
After both distances have converged, the system moves an infinitesimal time-step forward and the whole procedure is repeated having as initial guess for the distances the values calculated in the preceding time step. This convergence procedure is quite robust and is not too sensitive to small differences in the length of the time interval selected, or to the chosen initial values adopted for the distances. 

\subsection{Determination of the effective parameters entering the theory}
The  Langevin Equation for cooperative dynamics depends on a number of parameters that are determined either from the analysis of computer simulations or from experiments. These parameters represent well-defined physical quantities and are evaluated either by independent measurements or by theoretical calculations. Here we compare the outcome of the optimization procedure with data from the literature and show that the optimized parameters are consistent with other independent results.

The parameters include molecular parameters, and thermodynamic parameters. Both the non-Gaussian parameter, $\alpha_2$, and the semiflexibility parameter, $g$, are calculated from simulations. Other parameters entering the theory are the entanglement length, $d$, the number of correlated chains, $n$, and the monomer friction coefficient, $\zeta_{eff}$, which are obtained in this study through direct optimization of the theory by comparison with experiments.

The distance $d$ is equivalent to the statistical length of the chain segment between two entanglements, $d=N_e |\textbf{l}|$. The value of $N_e$ has been traditionally measured using different  experimental methods such as melt rheology, neutron spin echo, and NMR relaxation.  Unfortunately different methods give different values of $N_e$. This discrepancy  can result from the fact that different experimental methods explore different times scales. Additionally, there are possible artifacts in each type of measurement and there are differences in the theoretical models used to interpret the data and to calculate the entanglement degree of polymerization. For polyethylene the number of monomers between two entanglements has been estimated to be $N_e=130$ and, given that the carbon-carbon bond length is $l=1.53 \ \AA$, the estimated value of $d \approx 20 \ nm$.

Some recent numerical procedures have been designed to evaluate $N_e$ from an analysis of computer simulations such as the primitive path method\cite{PPA}, the Contour REduction Topological Analysis algorithm or CRETA algorithm,\cite{creta,creta1} and the Z-code.\cite{zcode1,zcode2} These methods give slightly different values of $N_e$ for polyethylene and consequently the entanglement distance $d$ is estimated to be somewhat smaller than from the experimental estimate. 
In our study the calculation of $d$ by direct comparison of the theory with the data of neutron spin echo gives identical outcome for all the entangled samples even if the parameter is evaluated for each sample independently and is consistent with $d=\sqrt{N_e} l$ for $l=1.53 \ \AA$ and $N_e\approx 130$ for polyethylene.\cite{marPRE}

Another input parameter is the monomer effective friction coefficient which is obtained in this study using two different procedures depending on the degree of polymerization of the samples. Short chains, which follow unentangled dynamics, have fast relaxation and reach the region of Fickian dynamics during the timescale of the NSE experiments. For these samples the monomer friction coefficient is calculated from the diffusion coefficient in the region for  $t> \tau_d$, where the chain mean-square-displacement scales linearly in time, as $\zeta_{eff}=k_BT/(N D)$, with $k_B$ the Boltzmann constant. The regime of Brownian motion is also sampled well in UA-MD simulations which provide complementary data to the experiments.

\begin{figure}
\includegraphics[width=3.5in]{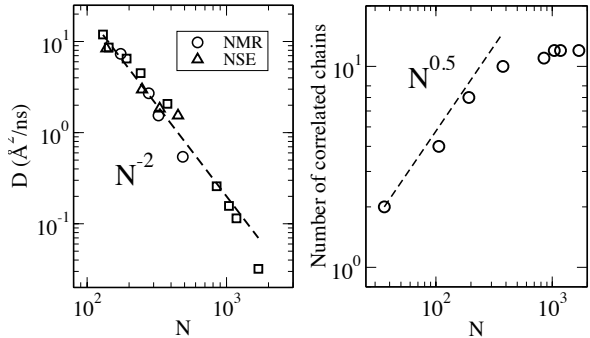} \\
\caption{Optimized parameters. Left panel: diffusion coefficient of entangled samples, calculated using the numerically optimized friction coefficient (squares) as a function of the degree of chain polymerization, and compared with NMR data from Ref. \cite{Pearson,Pearson1} (circles), and NSE data from Ref. \cite{Richter2} (triangle). Right panel: numerically optimized number of macromolecules undergoing cooperative dynamics as a function of the molecular degree of polymerization. Crossover to entangled dynamics is at $N_e=130$. Reprinted figure with permission from [M. G. Guenza Phys. Rev. E \textit{89}, 0526031 (2014).] Copyright (2014) by the American Physical Society.}
\label{figurexx}
\end{figure}

The friction coefficient of long entangled chains is more difficult to derive from experiments because those samples do not reach the diffusive regime, and the friction is calculated as a free parameter optimized from the direct comparison of the theory vs. experiments. However, the resulting diffusion coefficients, which are displayed in the left panel of Figure \ref{figurexx}, for the entangled polymers are consistent with the values obtained from independent exeriments of NMR \cite{Pearson,Pearson1} and NSE,\cite{Richter2} and follow the scaling with $N$ expected for long chains with this number of entanglements.\cite{Zamponi}
They are also consistent with the values of diffusion coefficients that we calculated using our reconstruction procedure from the mesoscale simulations of coarse-grained polyethylene melts.\cite{PRELiub,JCPLiub} 

The last non-trivial parameter is the number of chains undergoing correlated dynamics, $n$, which is given by the statistical number of polymers that occupies the volume defined by the range of the intermolecular potential. The range of the potential, for samples at the density and temperature of the experimental data, is of the order of the polymer radius of gyration, $R_g=\sqrt{N}l_{eff}/6$. Given that for melts, the monomer density $\rho=nN/R_g^3\approx 1$, the number of correlated chains is $n\approx \rho \sqrt{N} l_{eff}^3$. Therefore, when $N$ is fixed, $n$ is expected to increase with increasing stiffness and/or degree of polymerization, as the overall volume spanned by one chain increases, and with increasing density, as the 
number of monomers that fills the unit of volume increases. 
For chains of melts of homogeneous composition and constant density, the number of chains undergoing cooperative dynamics should grow as $ n \propto N^{0.5}$.\cite{Zamponi}

In this study the number of chains undergoing cooperative dynamics is evaluated as an adjustable parameter whose value is defined by optimizing the direct comparison of the theory to the experiments. The parameter is found to increase as $N^{0.5}$ for unentangled and slightly entangled chains, but to become fairly constant in the entangled regime(see right panel of Figure \ref{figurexx}).\cite{marPRE} The cooperative motion can grow until the crossover to entangled dynamics, where the cooperative dynamics becomes confined to the region between entanglements, or, to use a picture close to the ``reptation" model, in entangled polymers the cooperative motion, and the related sub-diffusive behavior, is dominated by the presence of entanglements, which confine the correlated motion of the chain to the region inside the ``tube".


\section{The dynamic structure factor and the alpha parameter}
\label{alphaparameter}

\begin{figure}
\includegraphics[width=3.5in]{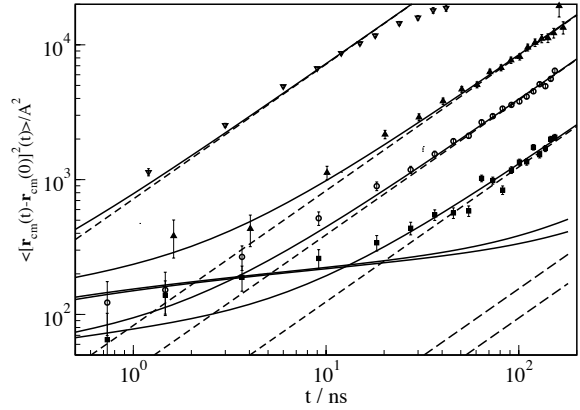} \\
\caption{Center-of-mass mean square displacement as a function of time for polyethylene melts of increasing chain length. Symbols are data from Neutron Spin Echo, lines are the theory presented here. From top to bottom, $N=36$ (riangle down), $106$ (triangle up), $192$ (circle), $377$ (squares), $N=1041$ and $N=1178$ (superimposing lines). Long-time Fickian dynamics is represented by the dashed lines.}
\label{msd}
\end{figure}

Figure \ref{msd} displays the center-of-mass mean-square displacement  as a function of time, including both the experimental values from Neutron Spin Echo and the theoretical predictions. For short chains the mean-square displacement is derived from the low $k$ values of the experimental dynamic structure factor, but for long entangled chains the values of momentum transfer $k$ of the NSE experiments are not low enough to isolate the center of mass dynamics and only the theoretical predictions are reported. 

At short time the motion of the center of mass is ballistic and at long time is diffusive. In the intermediate region the dynamics is sub-diffusive, with $\Delta R(t)^2 \propto t^{\nu}$ and $\nu< 1$, even for unentangled samples. In undercooled systems the subdiffusive behavior is a signature of the presence of heterogeneous dynamics.

 Polymers are glass forming systems and their dynamics is frustrated by the competing effects of the high degree of chain interpenetration and the presence of excluded volume interactions. We argue that this leads to the observed subdiffusive behavior, which is present in both unentangled and entangled systems, with the caveat that entanglements contribute to the extent of subdiffusive dynamics.
The sub-diffusive regime crosses over to diffusive at the characteristic decorrelation time, $\tau_d$, previously defined in this paper. For time intervals longer than $\tau_d$ the dynamics is Brownian and diffusive.

The extent that the mean-square displacement differs from Brownian motion can be quantified by calculating the alpha parameter, also called  the non-Gaussian parameter, defined for the monomer mean-square displacement as
\begin{eqnarray}
\alpha_2(t) = 3/5 <[\Delta \mathbf{r}(t)]^4>/<[\Delta \mathbf{r}(t)]^2>^2-1 \ ,
\end{eqnarray}
with $<[\Delta \mathbf{r}(t)]^2>= < [\mathbf{r}_{i}(t) - \mathbf{r}_{j}(0)]^2> $ the monomer displacement.
For a Gaussian distribution of displacements, which characterizes Markov statistics and so diffusive dynamics, the fourth moment of the distribution is trivially related to the square of the second moment as $<[\Delta \mathbf{r}(t)]^4>/<[\Delta \mathbf{r}(t)]^2>^2=5/3$ and $\alpha_2(t)=0$.

In principle, this can be calculated from the Fourier transform of the incoherent part of the dynamic structure factor, measured experimentally.  However the number and range of $k$ available in most studies are limited and do not consent a precise calculation by Fourier transform.\cite{colmenero1} A different strategy is to investigate the non-Gaussian nature of the Van Hove function from the direct analysis of simulations of the polymer melt performed in the same thermodynamic conditions of the experimental data.  
We have calculated the alpha parameter from simulations we performed at the same thermodynamic conditions of the experimental data, for polyethylene at the temperature $T=509 \ K$, and density $\rho=0.733$ $g/cm^3$, and increasing degree of polymerization.
The melts of polyethylene chains have increasing degree of polymerization, $N=36$  (number of chains $=350$), $44 \ (350)$, $66 \ (350)$, $78 \ (350)$, $100 \ (350)$, $192 \ (350)$, $224 \ (32)$, and $270 \ (32)$, where we report in parenthesis for each sample the number of chains included in the simulation box. 
Simulations were performed with LAMMPS code\cite{LAMMPS} and run in parallel using the SDSC Trestles cluster available through XSEDE.\cite{XSEDE} All the simulations were performed in the canonical ensemble using the Nose-Hoover thermostat.\cite{ClarkPRL,Jay2} 

The experimental data we analyze are from NSE measurements of the incoherent part of the intermediate scattering function, which is defined as 
\begin{eqnarray}
S(\textbf{k},t)= N^{-1} \sum_{i,j=1}^N <e^{i \mathbf{k} \cdot [\mathbf{r}_i(t) - \mathbf{r}_j(0)]}> \ .
\end{eqnarray}
\noindent $S(\textbf{k},t)$ can be written as the exponential function of an infinite series in the momentum transfer $k^2$.\cite{Kubo,Sjolander} In the condition under which the distribution of displacement is Gaussian only the $k^2$ term is relevant, while for non-Gaussian distributions higher order corrections need to be included.\cite{Zorn} The dominant term in the non-Gaussian correction is of order $k^4$,
\begin{eqnarray}
<e^{i \mathbf{k} \cdot \Delta \mathbf{r}(t)}> \approx e^{- \frac{k^2}{6}<[\Delta \mathbf{r}(t)]^2> + \frac{k^4}{72}<[\Delta \mathbf{r}(t)]^2>^2  \alpha_2(t) - O(k^6)} \ .
\label{alpha}
\end{eqnarray}
For a system that obeys Gaussian statistics $\alpha_2(t)=0$, so that for polymer melts at the thermodynamic conditions of this study $\alpha_2(t)=0$ in the ballistic regime, which is not present in the NSE measurements, and in the diffusive regime ($t >> \tau_d$). In the intermediate regime $\alpha_2(t) \neq 0$ and this contribution has to be accounted for in the calculation of the dynamic structure factor.

The systems simulated show the presence of non-Gaussian displacement distributions in the region of sub diffusive dynamics. The samples are largely in the unentangled regime, so that the correction to include in the dynamic structure factor for the highly entangled chains cannot be quantitatively determined in our simulation. However, the contribution for entangled systems is calculated by extrapolating the unentangled behavior.\cite{marPRE} 

\section{Comparison of the theory with experiments of dynamic structure factor}
\label{comparison}
We compare the prediction of the theory for the dynamic structure factor, with Neutron Spin Echo data for polyethylene samples at the temperature $T=509 \ K$, and density $\rho=0.733$ $g/cm^3$.\cite{richter1} The bond length of the experimental samples is $1.53 \ \AA$, while the effective bond length, obtained by imposing that statistical segments are intramolecularly uncorrelated, is $l_{eff}^2=17 \ \AA^2$. This length corresponds to an effective semiflexibility parameter of $g=0.785$, calculated by optimization of the chain structural properties such as the mean square end-to-end distance in comparison with simulations.\cite{perslength,booth}

We calculate the incoherent scattering function for a dynamically heterogeneous polymer system, by including the $q^4$ correction term to the Gaussian approximation, as described in Eq.(\ref{alpha}).  The agreement between theory and experiments is quantitative for every system.\cite{Zamponi, marPRE}

This comparison for unentangled samples was already discussed in a preceding paper, where the theoretical predictions for an extended Rouse approach with cooperative dynamics and chain semi-flexibility showed quantitative agreement with the NSE data of the intermediate scattering function.\cite{Zamponi} The theory in its present form produces data that are indistinguishable from the results from its previous version when it is applied to the study of  unentangled chains. Here the presence of entanglements does not affect the dynamics because the monomer decorrelation occurs at a timescale shorter than the time the monomers inter-diffuse an average distance comparable with $d$.

\begin{figure}
\includegraphics[width=3.0in]{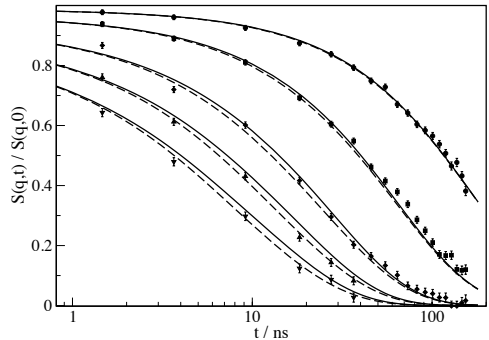} \\
\caption{Comparison between  theoretical (lines) and experimental (symbols) values of the normalized incoherent intermediate scattering function for polyethylene with $N=192$ with (full lines) and without (dashed lines) the contribution due to entanglements. Data are 
at increasing wave vector $q=0.3$ (circle), $0.5$ (square), $0.77$ (diamond), $0.96$ (triangle up), $1.15$ (triangle down). }
\label{figura6}
\end{figure}

\begin{figure}
\includegraphics[width=3.0in]{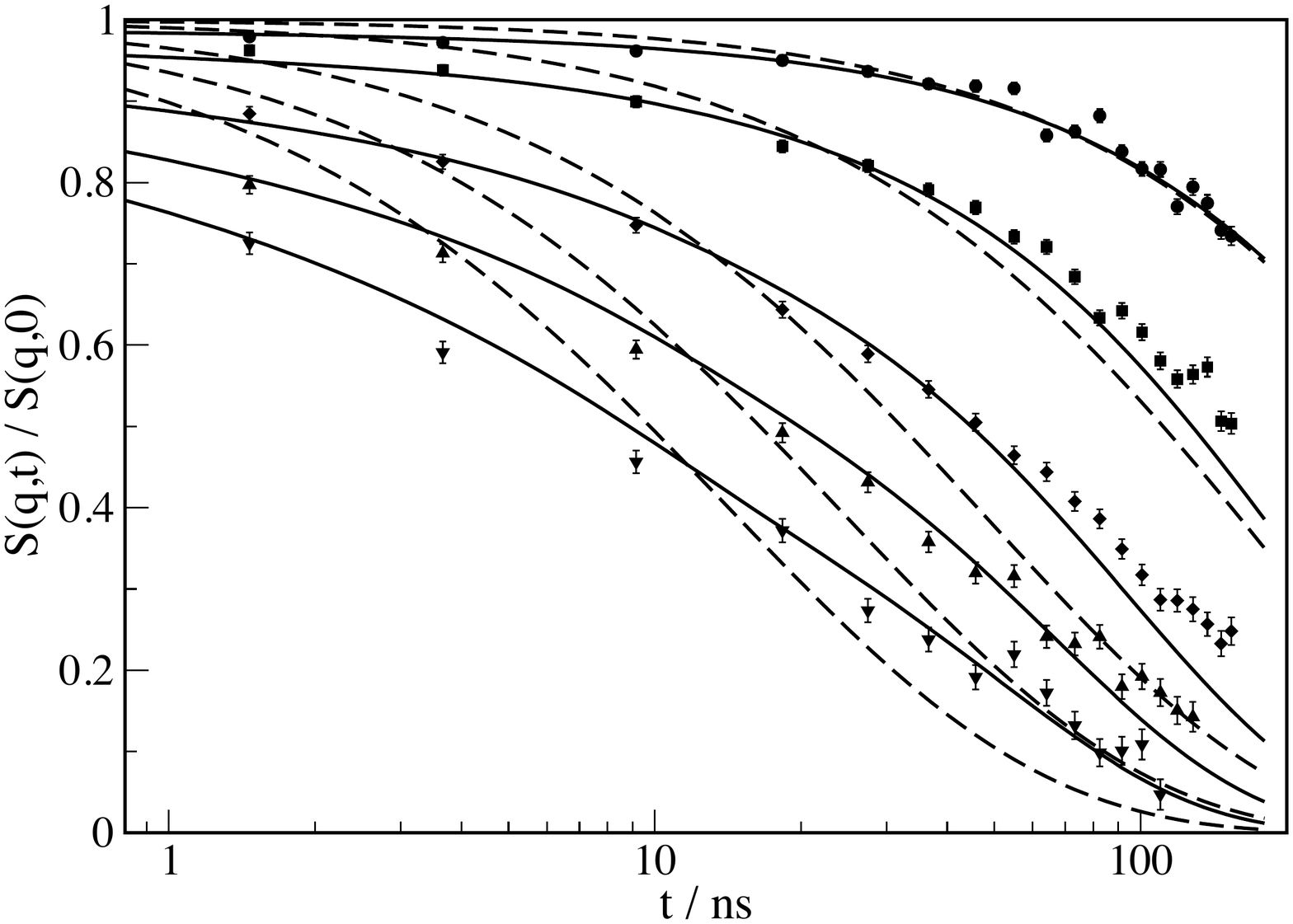} \\
\caption{Comparison between  theoretical (lines) and experimental (symbols) values of the normalized incoherent intermediate scattering function for polyethylene with $N=377$ with (full lines) and without (dashed lines) the contribution due to entanglements. Data are 
at increasing wave vector $q=0.3$ (circle), $0.5$ (square), $0.77$ (diamond), $0.96$ (triangle up), $1.15$ (triangle down). }
\label{figura7}
\end{figure}

\begin{figure}
\includegraphics[width=3.0in]{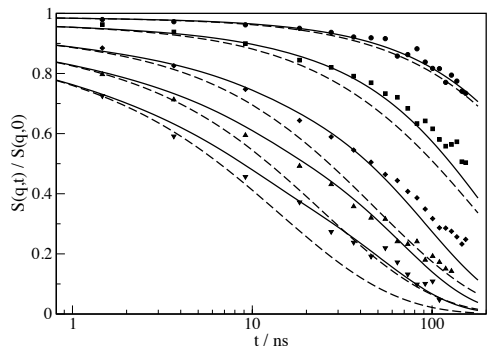} \\
\caption{Comparison between  theoretical (lines) and experimental (symbols) values of the normalized incoherent intermediate scattering function for polyethylene with $N=377$ with entanglement (full lines) and the Rouse dynamics of a semiflexible chain without entanglements and without cooperative dynamics (dashed lines). Data are 
at increasing wave vector $q=0.3$ (circle), $0.5$ (square), $0.77$ (diamond), $0.96$ (triangle up), $1.15$ (triangle down). }
\label{figura8}
\end{figure}

\begin{figure}
\includegraphics[width=3.0in]{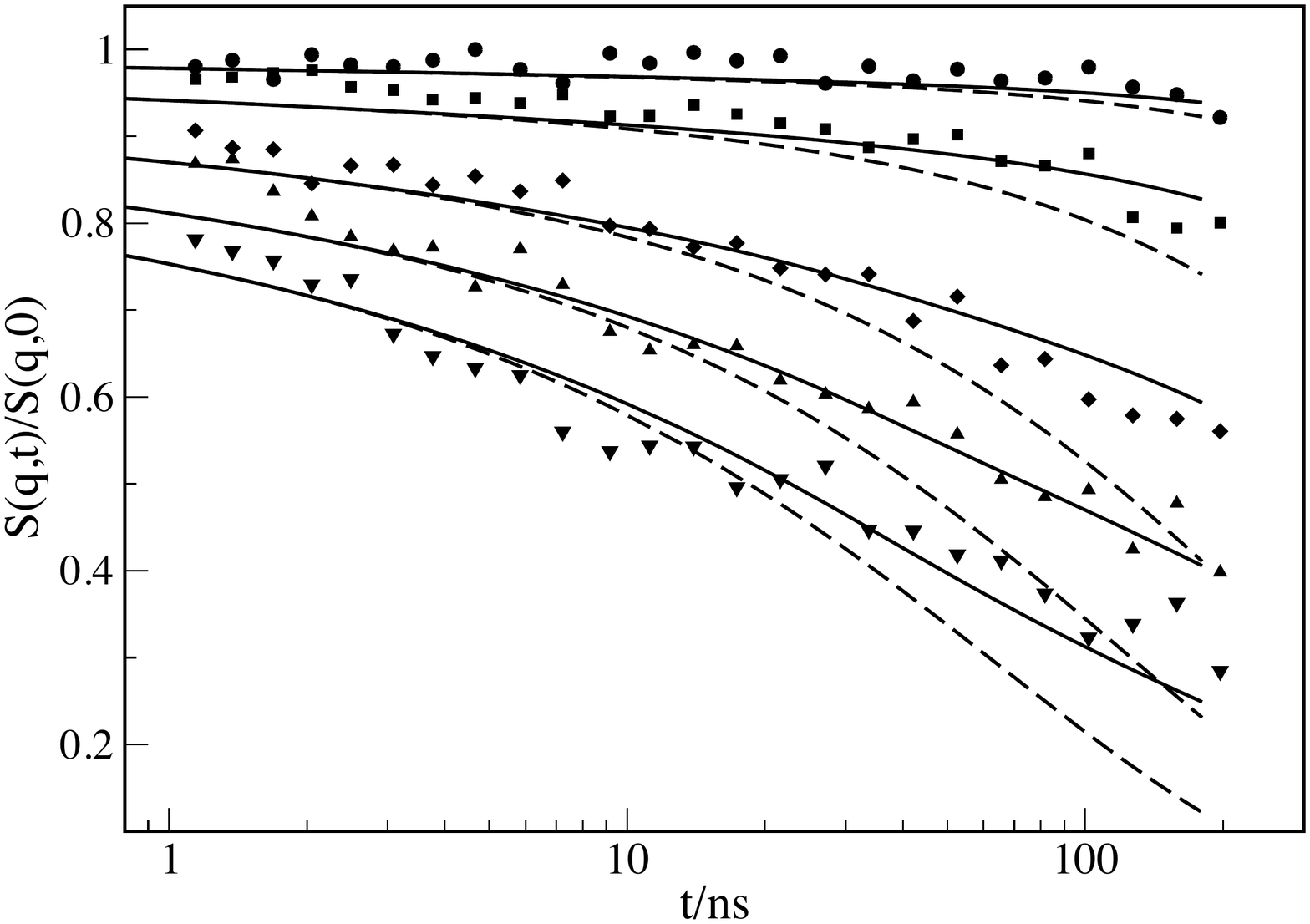} \\
\caption{Comparison between  theoretical (lines) and experimental (symbols) values of the normalized incoherent intermediate scattering function for polyethylene with $N=1041$ with (full lines) and without (dashed lines) the contribution due to entanglements. Data are 
at increasing wave vector $q=0.3$ (circle), $0.5$ (square), $0.77$ (diamond), $0.96$ (triangle up), $1.15$ (triangle down). }
\label{figura9}
\end{figure}

\begin{figure}
\includegraphics[width=3.0in]{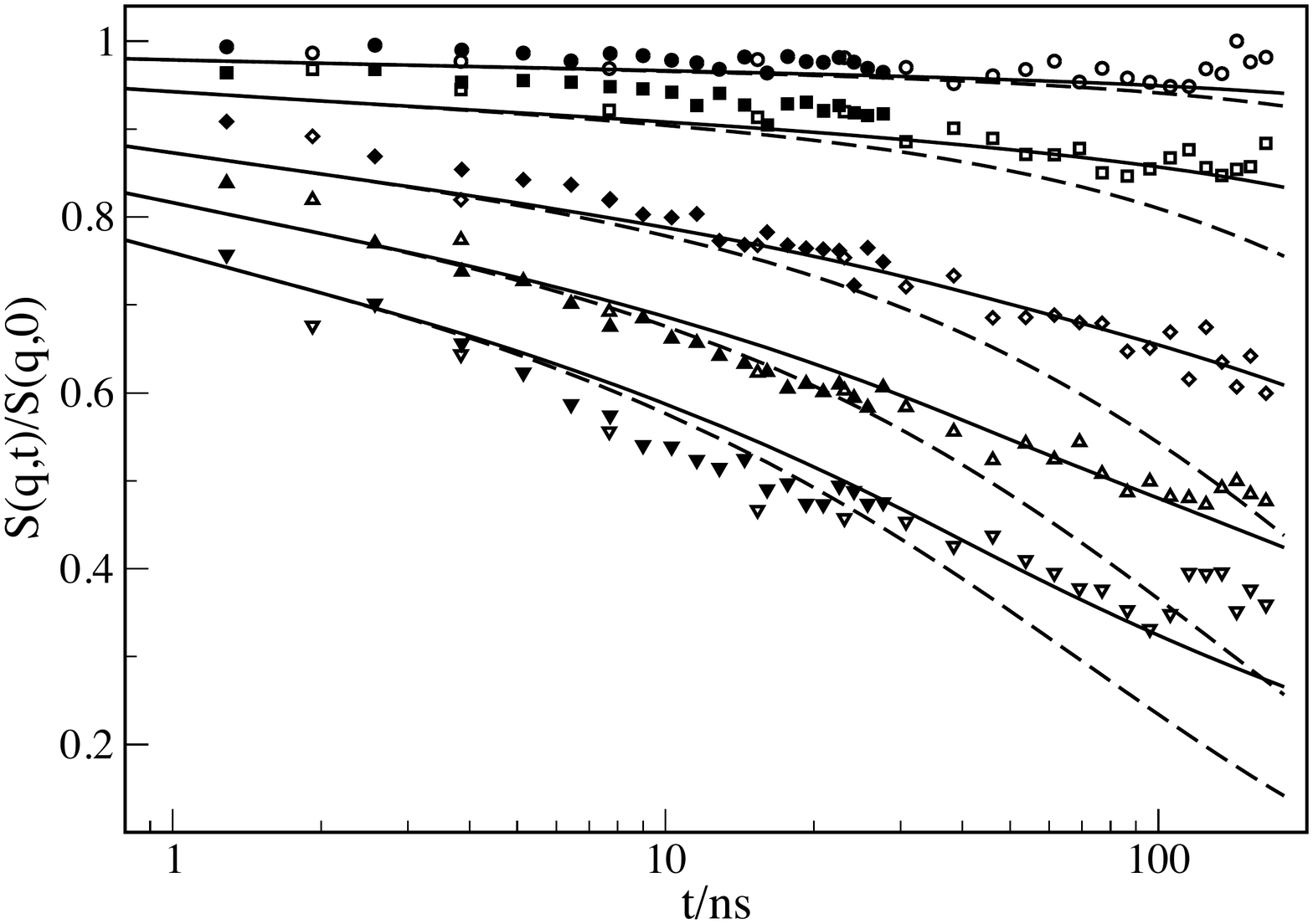} \\
\caption{Comparison between  theoretical (lines) and experimental (symbols) values of the normalized incoherent intermediate scattering function for polyethylene with $N=1178$ with (full lines) and without (dashed lines) the contribution due to entanglements. Data are 
at increasing wave vector $q=0.3$ (circle), $0.5$ (square), $0.77$ (diamond), $0.96$ (triangle up), $1.15$ (triangle down). }
\label{figura10}
\end{figure}

\begin{figure}
\includegraphics[width=3.0in]{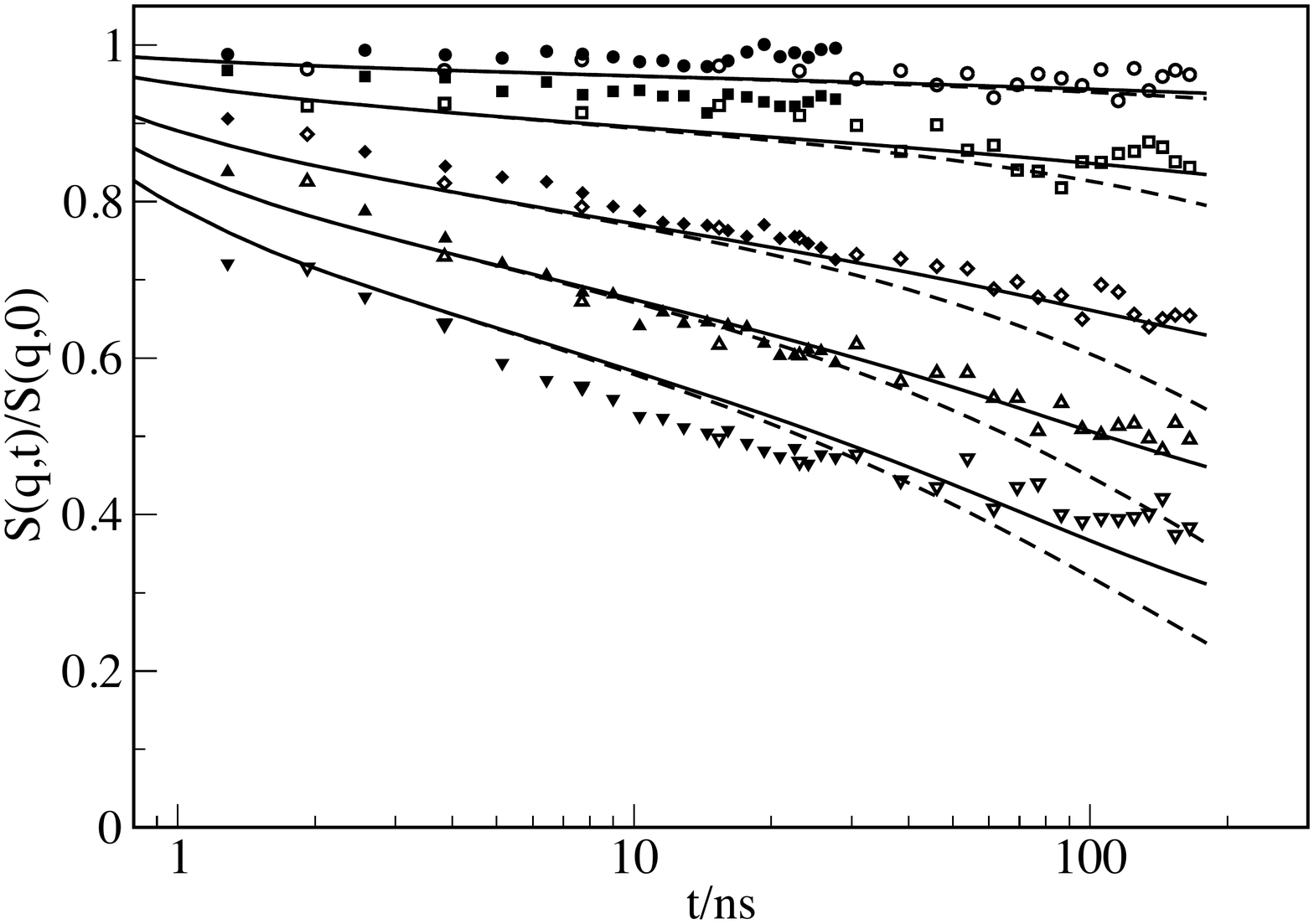} \\
\caption{Comparison between  theoretical (lines) and experimental (symbols) values of the normalized incoherent intermediate scattering function for polyethylene with $N=1692$ with (full lines) and without (dashed lines) the contribution due to entanglements. Data are 
at increasing wave vector $q=0.3$ (circle), $0.5$ (square), $0.77$ (diamond), $0.96$ (triangle up), $1.15$ (triangle down). }
\label{figura11}
\end{figure}

First we report here the analysis of the polyethylene melt with $N=192$. This chain has statistically only one entanglement, and its dynamics is mostly well described by the Langevin Equation for cooperative dynamics with no entanglements. In Figure \ref{figura6} experimental data are compared with the theory, where the contribution from entanglements is accounted for, and with the theory where entanglements are neglected. The difference between the two curves is small but the correction due to the presence of the entanglements still improves the agreement of the theory with the data.

As a second example, the theory is compared with NSE data for a sample with $N=377$. Each chain in the melt has statistically two entanglements. Thus this sample is in the crossover region between unentangled and entangled dynamics. Fig. \ref{figura7} displays the comparison between the theoretical predictions of the Langevin Equation for the cooperative dynamics with monomer constraint potential. The figure also displays the predictions of the Rouse model for a semiflexible polyethylene chain of identical chain length, semiflexibility, and monomer friction coefficient. The second curve represents the calculation of the dynamics without including cooperative many-chain motion, nor the constraint due to entanglements. This curve is not only quantitatively but also qualitatively in disagreement with the data, indicating that the description of a semi flexible chain with a flexible model gives an incorrect representation of the dynamics (unless the statistical segment is chosen larger than the persistence length of the polymer).

In the following figure (Fig. \ref{figura8}) experimental data for $N=1041$ are compared with the theory where semiflexibility and cooperativity of the dynamics, but not entanglements, are accounted for. The theory agrees well with the experiments in the short time region, for $t \le 2 ns$ as the chain in the initial regime has not yet experienced the presence of entanglements, but does not reproduce the data in the long-time regime. In the long-time regime it is essential to include the contribution from entanglement to recover the experimental data with quantitative agreement.

Similar plots are presented for chains with a high number of entanglements, namely $N=1178$ corresponding to $9$ entanglements and $N=1692$ corresponding to $13$ entanglements. Those plots are shown in Figure \ref{figura9} and in Figure \ref{figura10}, respectively. One of the  advantages of this theory is that it is possible to analyze separately the different contributions to the polymer dynamics.

In all the plots the theory without the presence of entanglements agrees well with the experiments in the short time region, for $t \le 2-3 ns$. This general behavior appears to be reasonable considering the fact that at short time the monomer dynamics is not affected by the length of the polymer chain the monomer belong to, because at short time the monomer only samples its local environment. A common local dynamics is in agreement with the  physical hypothesis that motivates the coupling model approach, where  local and global dynamics connect at an intermediate time scale which is similar for all polymers.\cite{Kia}

\section{Conclusions}
This paper present a theoretical approach to the dynamics of polymer melts covering systems with different degrees of polymerization, spanning the dynamics from the unentangled to the fully entangled regime. The theory doesn't need to be modified to describe the two different dynamical regimes observed for short and for long chains,  and, to the best of our knowledge, it represents the first unified approach to polymer dynamics across the transition from unentangled to entangled motion.

The flexibility of the formalism in addressing different systems is due to the many-chain nature of the approach, which allows for the simultaneous description of the dynamics of a group of interpenetrating polymer chains that are interacting and, if the chains are long enough, are also mutually entangled. As polymers are fractal object of order two, they expand in volume with increasing degree of polymerization, faster than they can fill their own volume. In this way the number of interpenetrating chains grows with chain length as $N^{1/2}$. The dynamics in  the entangled regimes reflects the interplay of the cooperative dynamics of  $N^{1/2}$ macromolecules and entanglements.

The theory selects at initial time a group of interpenetrating chains and monitors at increasing time intervals their dynamics as the chains interdiffuse and their motion mutually decorrelates. The chains are interacting through an effective center-of-mass pair potential that represents the projection of the many-body monomer-monomer interactions onto the polymer center-of-mass of a pair of chains. The monomer-monomer interactions propagate through the liquid of macromolecules surrounding the slow-moving chains undergoing cooperative motion. The mean field contribution due to the units that are projected out during the coarse-graining of the dynamics, which leads to the Langevin equation, result in an effective potential which is long-ranged and couples the dynamics of the slow moving chains.

A second intermolecular potential that is included in the theory at the monomer level is the direct consequence of the chains being entangled. This potential limits the relative motion of two monomers belonging to two different chains, which are initially in contact and then interdiffuse freely until they experience the constraint in their dynamics due to entanglements. Both potential depend in magnitude on the distances between the interacting units, which is evolving in time as the chains interdiffuse. In this way the system is considered in equilibrium only locally in the time domain, and the acting forces are solved self-consistently at any given time interval. Finally all the molecules that are initially correlated become uncorrelated and at long enough time intervals the sampled dynamics is Brownian. In this long time window, however, the effect of the entanglements is still present and affects the diffusion. The latter reproduces quantitatively the experimental data in both the unentangled and entangled regimes. 

Chains that are initially interpenetrating need to move in a cooperative way for their dynamics to becomes both inter- and intramolecular uncorrelated. The dynamical mechanism of relaxation is conducive to the relaxation mechanisms observed in undercooled polymer liquids approaching their glass transition, with slow rearranging regions and fast motion of a few chains. 

The chain dynamics is characterized by the interplay between cooperative dynamics and the presence of entanglements. The latter tend to reinforce cooperative motion for the monomers that are comprised in the volume defined by the chain segment between two entanglements. In this way we observe that the region of cooperative dynamics increases in volume with the square of the chain length until polymers reach a length comparable with the entanglement lengthscale. At that point the region of cooperative motion is the same for all the samples with larger chain length that are analyzed in this study, and it is equal to the lengthscale of the entanglement length. It is reasonable to hypothesize that for chains longer than the ones described in this study, a secondary correlated motion could occur on the longer lengthscale of the chain size where the cooperativity would involve the domains on the entanglement lengthscale in a hierarchical clustering of cooperative motion.

\section{Acknowledgements}
The author  thanks Michal S. Jander for the careful reading of the manuscript.
This material is based upon work partially supported by the National Science Foundation under Grant No. DMR-0804145 and CHE-1362500.
This work used the Extreme Science and Engineering Discovery Environment (XSEDE), which is supported by National Science Foundation grant number ACI-1053575.

\end{document}